\documentclass[useAMS,usenatbib]{mn2e}

\usepackage{graphicx}
\usepackage{subfigure}
\usepackage{color}
\usepackage[authoryear]{natbib}
\usepackage{names}

\begin{document}
\title[Post-main sequence evolution of A star debris discs]{Post-main sequence evolution of A star debris discs}

\author[A. Bonsor et al.]
  {A. Bonsor$^1$\thanks{Email: abonsor@ast.cam.ac.uk},
   M. C. Wyatt$^1$\\
  $^1$ Institute of Astronomy, University of Cambridge, Madingley Road,
  Cambridge CB3 0HA, UK}

\maketitle

\begin{abstract}

 While the population of main sequence debris discs is well constrained, little is known about debris discs around evolved stars. This paper provides a theoretical framework considering the effects of stellar evolution on debris discs, particularly the production and loss of dust within them. Here we repeat a steady state model fit to disc evolution statistics for main sequence A stars, this time using realistic grain optical properties, then evolve that population to consider its detectability at later epochs. Our model predicts that debris discs around giant stars are harder to detect than on the main sequence because radiation pressure is more effective at removing small dust around higher luminosity stars. Just 12\% of first ascent giants within 100pc are predicted to have discs detectable with Herschel at 160$\mu$m. However this is subject to the uncertain effect of sublimation on the disc, which we propose can thus be constrained with such observations. Our model also finds that the rapid decline in stellar luminosity results in only very young white dwarfs having luminous discs. As such systems are on average at larger distances they are hard to detect, but we predict that the stellar parameters most likely to yield a disc detection are a white dwarf at 200pc with cooling age of 0.1Myr, in line with observations of the Helix Nebula. Our model does not predict close-in ($<$0.01AU) dust, as observed for some white dwarfs, however we find that stellar wind drag leaves significant mass ($\sim$10$^{-2}$M$_{\oplus}$), in bodies up to $\sim$10m in diameter, inside the disc at the end of the AGB phase which may replenish these discs.

\end{abstract}


\section{Introduction}

The first dusty disc around a main sequence star was observed in 1984 around Vega \citep{aumann1984}. Since then our knowledge of such systems has improved significantly, and it is now known that 32\% of A stars exhibit excess emission in the infrared, over and above the stellar photosphere \citep{su06}. This is thermal emission from dust particles orbiting the star in a debris disc. Debris discs are collisionally dominated in that the smallest bodies in the system are continuously replenished by collisions between larger objects and are subsequently removed by radiation pressure.  The long term evolution of such systems can be modelled by steady state collisional models \citep{krivov08, wyatt07, KB04, grantstirring}. Disks are depleted due to collisional erosion and are expected to show a slow decline in brightness. A decrease in brightness with age is indeed observed i.e \citep{su06} and can be well fitted with the models of \cite{wyatt07}, allowing such models to characterize the population of main sequence A stars debris discs reasonably accurately. These models assume that velocities in the disc are high enough that collisions are destructive. This requires that the disc is stirred, for exmple by self-stirring (e.g. \cite{KB04}) or planet stirring (e.g. \cite{alex}.

Dust is also seen around some post-main sequence stars. In some cases this dust can be a result of the evolution of the star, for example material emitted in the stellar wind form spherical shells of dust that are observed around AGB stars (e.g. \cite{AGBshell2010}) or even stable discs observed around post-AGB stars, possibly linked to binarity (e.g.~\cite{winkel09postAGBbinary}). Infrared excess observed around giant stars, e.g. \cite{jura99}, and the helix nebula \citep{helix}, on the other hand, has been interpreted as a disc similar to debris discs on the main sequence (although alternative interpretations do exist see e.g. \cite{kimzuckerman01}). Hot dust is also observed in small radii ($<$0.01AU) discs around white dwarfs, e.g. \cite{farihi09} or \cite{2007Kuchner}, again inferred to originate from a debris disc. However, in contrast to main sequence debris discs, these discs cannot be in steady state since material at such small radii has a short lifetime. Rather models suggest that these discs are formed when an asteroid approaches close to the star where it is tidally disrupted \citep{jurasmallasteroid}.

There are not yet enough observations of discs around post-main sequence stars to fully understand the population and it is not clear how the few discs that have been discovered around post-main sequence stars relate to the progenitor population of debris discs on the main sequence. In this work we take advantage of the fact that the main sequence debris disc population around A stars is well characterised and extend the steady state collisional evolution models to consider the changes to this known population during the star's evolution. In particular we consider its detectability on the post-main sequence and therefore whether the observed post-main sequence discs derive from this population and what future observations would be best suited to detect them. 

 Previous work has looked at specific aspects of the effects of stellar evolution on asteroids or comets, such as stellar wind drag \citep{dong10} and sublimation \citep{juraotherkb, jurasmallasteroid}. \cite{dong10} model the evolution of a planetesimal belt due to stellar mass loss and suggest that the captue of km-sized planetesimals into mean motion resonance could explain systems such as the helix nebula. In this work a theoretical framework is built that incorporates the effect of collisions, radiation forces, the stellar wind, sublimation and realistic optical properties of dust, during the star's evolution from the main sequence to the white dwarf phase, focussing on the observable properties of the belt. The dynamical effects of stellar evolution, in particular stellar mass loss, on planetary systems will be considered in future work. 

This paper begins by discussing the evolution of the star in \S \ref{sec:star}. In \S \ref{sec:modelmw} we revisit the initial population of debris discs around main sequence A stars this time reproducing the fit to Spitzer observations at 24 and 70$\mu$m of \cite{wyatt07} using a model that incorporates the optical properties of realistic grains rather than a black body approximation. The steady state collisional models of \cite{wyatt07} are then extended to include post-main sequence stellar evolution. In \S \ref{sec:models} the changes to the properties of an individual disc, as the star evolves, are described, whilst \S \ref{sec:obs} discusses the implications for observations of the population of debris discs around evolved stars, focusing in particular on giant stars and white dwarfs and \S \ref{sec:conclusions} summarises the models discussed.

\section[stellarevol]{Stellar Evolution}
\label{sec:star}

\begin{figure}
\includegraphics [width=80mm]{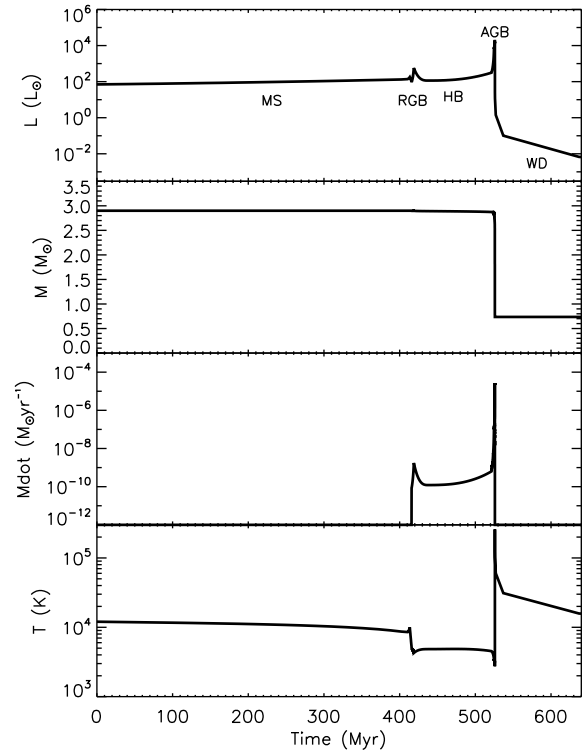}
\caption{The luminosity (L), mass (M), mass loss rate and temperature (T) evolution of a 2.9$M_{\odot}$ star, Z=0.02, in solar values. MS: main sequence (0-413Myr), RGB: red giant (415.9-418.7 Myr), HB: horizontal branch (418.7-521.4)Myr, AGB: asymptotic giant branch (521.4-525.9)Myr, WD: white dwarf (525.9 Myr onwards). }
\label{fig:star}
\end{figure}

Knowledge of the evolution of the star is required in order to model the post-main sequence evolution of the debris disc. The two crucial parameters that determine the disc's evolution and detectability are the stellar luminosity and mass. In this work stellar evolution models from \cite{SSE} are used. These models use analytic formulae and are designed for population synthesis. They are accurate to within 5\% of more detailed models. 

\par
Here we consider stellar evolution models for stars of mass 1.67 to 3.8 $M_{\odot}$, these correspond to stars of spectral type A9-B8, according to the models of \cite{kuruczstarparams}, although the stellar temperature varies along the main sequence and a star of a given mass will be classified differently depending on the point in its evolution at which it is observed. 

  The evolution of A stars can be split up into various phases (see Fig.~\ref{fig:star} for the evolution of an 2.9$M_{\odot}$ star, equivalent to spectral type A0). Along the main sequence (MS) the star is stable and powered by core hydrogen burning. There is a slight increase in luminosity, as the size of the hydrogen burning core increases, but mass loss is negligible. Main sequence luminosities, for the spectral type stars considered, range from $7.5 L_{\odot}$ (1.67$M_{\odot}$) to 180$L_{\odot}$ (3.8$M_{\odot}$) and main sequence lifetimes from 200Myr (3.8$M_{\odot}$) to 2000Myr (1.67$M_{\odot}$). Changes in stellar properties along the main sequence are negligible and therefore are ignored in \cite{wyatt07} and \S \ref{sec:modelmw}. Once hydrogen in the core is exhausted, hydrogen shell burning commences and the star swells to become a red giant (RGB). Its luminosity increases by several orders of magnitude. The RGB phase lasts for $\sim$10Myr, until core temperatures are hot enough that helium burning starts, possibly degenerately in a helium flash (for $M\leq 2.25M_{\odot}$), after which its luminosity decreases. For those stars where helium burning starts degenerately, the star reaches a higher luminosity on the red giant branch than for higher mass stars. This means that although the initial luminosity on the giant branch increases with stellar mass, the value at the tip of the giant branch has a minimum for stars that no longer start helium burning degnerately ($M\geq 2.25M_{\odot}$), increasing up to $\sim10^4 L_{\odot}$ for 1.67M$_{\odot}$ stars.
 The star's luminosity increases slowly as it moves along the Horizontal Branch (HB), which lasts for $\sim$100 Myr. Once Helium in the core is exhausted the star swells to become an Asymptotic Giant Branch (AGB) star and its luminosity increases yet further. Luminosities of evolved A stars can reach up to $2 \times 10^{4} L_{\odot}$ for a 3.8$M_{\odot}$ star in this phase. 
\par The majority of mass loss occurs on the AGB. There is a fair degree of uncertainty in mass loss rates and the exact mechanism driving the mass loss in AGB stars (see review by \cite{willson}). The stellar evolution code uses the mass loss of \cite{reimers78} on the red giant branch and horizontal branch:
\begin{equation}
\dot{M_R}=2\times 10^{-13} \frac{L_* R_* }{M_*} \; \; \mathrm{M_\odot yr^{-1}}
\end{equation}
where $M_*$, $R_*$ and $L_*$ are the stellar mass, radius and luminosity in solar values. 
\par  On the AGB the formulation of  \cite{VW} is used:
\begin{equation}
\label{mloss}
\log \dot{M}_{VW}= -11.4 + 0.0125\,[P -100\; \max(M_*-2.5,0.0)] \; \; \mathrm{M_\odot yr^{-1}}
\end{equation}
\begin{equation}
P=\min\,(3.3,-2.07-0.9\log M_*+1.94\log R_*)
\end{equation}
where P is the Mira pulsation period of the star, in days. The expansion velocity for the stellar wind is given by:
\begin{equation}
\label{vsw}
v_{SW}= -13.5 + 0.056 P \;\; \mathrm{kms}^{-1}.
\end{equation}

\par
 The mass loss rate peaks in a superwind phase, at the tip of the AGB \citep{VW}, with 
\begin{equation}
 \dot{M} = 1.36 \times 10^{-9} L_* \; \; \mathrm{ M_\odot yr^{-1}}.
\end{equation}
 These are empirical mass loss rates, fitted to observations of RGB and AGB stars. Thermal pulses, with periods $\sim10^5yr$ dominate the evolution on the AGB as the star switches between helium and hydrogen shell burning. This may lead to discrete superwind phases and the multiple shells seen in many planetary nebulae.

\par The white dwarf core now evolves at constant luminosity to higher effective temperature. This luminosity is proportional to the white dwarf mass. The hotter radiation from the white dwarf core ionizes the expelled gas which is seen as a planetary nebula. Once the white dwarf core reaches a maximum effective temperature it starts cooling at constant radius. The radiative cooling of the white dwarf is modeled by Mestel theory \citep{Mestel}, with the white dwarf luminosity, in solar units, given by \begin{equation}
L_{WD} =\frac{635 M_{WD} Z^{0.4}}{[A(t+0.1)]^\alpha}
\label{LWD}
\end{equation}
where $M_{WD}$ is the mass of the white dwarf ($M_{\odot}$), Z is the metallicity, t is the cooling age or time since the WD formed, in Myr and A is a parameter that is composition dependent. 
In the current models solar metallicity, Z=0.02 and A=15 for a CO white dwarf is used. Once the white dwarf has cooled significantly, crystallization occurs, and the cooling rate enters a different phase, and hence in these models $\alpha$ changes from 1.18 to 6.48 at an age of 9Gyr, for all spectral types.  Although prescriptions for white dwarf crystallization have improved significantly since the groundbreaking work of \cite{Mestel}, for example \cite{Met04}. However crystalisation only occurs when the white dwarf cools to $\sim 6000-8000$ K \citep{Met04} and therefore \cite{Mestel} should provide accurate luminosities for white dwarfs hotter than this. It is found later in this work that debris discs are only detectable around very young, hot white dwarfs (see \S ~\ref{sec:wdobs}) and therefore differences between the cooling theory of \cite{Mestel} and more modern prescriptions are not significant for the current work.

\section{Steady state models for population of debris discs around main sequence A stars}
\label{sec:modelmw}

\subsection{Evolution of an individual disc}
 The evolution of an individual disc, due to collisions, can be described by a simple steady state model such as that presented in \cite{wyatt07}. These models assume that the size distribution of all particles is the same as for an infinite collisional cascade:
\begin{equation}
n(D)dD \propto D^{-7/2}dD.
\label{eq:sizedistrib}
\end{equation}

This size distribution holds from the largest planetesimals of size, $D_C$, down to the smallest dust grains present of size, $D_{min}$ and assumes that all bodies of all sizes have the same strength per unit mass. In the steady state models for main sequence stars $D_{min}$ is determined by the largest grains that are not blown out of the system by radiation pressure, the blow-out size ($D_{\mathrm{bl}}$). 
 The blow-out size is an important property of the system and depends on the stellar parameters. Assuming spherically symmetric black body grains, of uniform density $\rho$ in kgm$^{-3}$, it is given, in $\mu$m by \citep{burns}:

\begin{equation}
\label{blow}
D_{\mathrm{bl}}= 0.8 \left(\frac{L_*}{M_*}\right) \left( \frac{2700}{\rho}\right ),
\label{eq:dbl}
\end{equation}
where $L_*$ and $M_*$ are the stellar luminosity and mass in solar units.
 
\par For such a cascade the majority of the cross-sectional area of the disc is found in the smallest particles, but the mass of the disc is determined by the largest objects. In these models the evolution of the mass in the disc ($M_{\mathrm{tot}}$ in $M_{\oplus}$) is determined by the collisional lifetime ($t_c$) of the largest objects, with diameter, $D_c$ in km:
\begin{equation}
\frac{dM_{\mathrm{tot}}}{dt}=\frac{-M_{\mathrm{tot}}}{t_c(D_c)}.
\label{eq:dmtot}
\end{equation} 

The collisional lifetime of particles of diameter $D_c$ was determined in \cite{wyatt07} and is given by: 
\begin{equation}
t_{c} =5.20\times 10^{-13}\frac{\rho \; r^{13/3} (\frac{dr}{r}) Q_D^{* 5/6}D_{c}}   {M_{\mathrm{tot}}M_*^{4/3}e^{5/3}} \;\;\; \mathrm{Myr}
\label{eq:tc}
\end{equation}

where r is the radius of the disc, in AU, and dr is the width of the disc, taken to be $\frac{r}{2}$, e is the eccentricity of the particles, assumed to be equal to their inclination I, $Q_D^*$ is the dispersal threshold for collisions in Jkg$^{-1}$ and $\rho$ is the density of particles, taken to be 2700kgm$^{-3}$. So long as the only time dependent variable in Eq.~\ref{eq:dmtot} is the mass in the disc, it can be shown that the mass in the disc evolves as:
\begin{equation}
\label{M}
M_{\mathrm{tot}}(t)=\frac{M_{\mathrm{tot}}(0)}{(1+\frac{t}{t_c})}.
\end{equation}
 For early times, when $t<<t_c$, the mass in the disc remains approximately constant at its initial value, only turning over and falling off as $M_{\mathrm{tot}} \propto \frac{1}{t}$ at times $t>>t_c$. At late times (t in Myr) the mass in the disc tends to a value $M_{max}$ that is independent of its initial value,
\begin{equation}
\label{eq:mmax}
M_{max} (t)= 5.2\times 10^{-13} \frac{\rho \; r^{13/3} (\frac{dr}{r}) Q_D^{* 5/6}D_{c}}{M_*^{4/3}e^{5/3} t} \;\;\; M_{\oplus}.
\end{equation}

These models are, however, a simplification. It is only considered that collisions between the largest bodies change the mass in the disc. Only a single value for the dispersal threshold, $Q^*_D$, is used and cratering collisions (e.g ~\cite{Kobayashi}) are ignored. However this simple procedure allows us to fit the observations. The calculated dust luminosities are within an order of magnitude of more detailed prescriptions in which collisions between all diameter particles are considered, $Q^*_D$ is a function of diameter and the size distribution is three-phase, for example \citep{lohne}. 

\par In \cite{wyatt07} the emission properties of the disc were calculated using a black body approximation. For reasons that are explained in \S ~\ref{sec:radp}, a black body approximation cannot be used in the current models. The models of \cite{wyatt07} were updated to incorporate realistic emission properties of grains using the method of \cite{wyattdent02}. The prescription of \cite{ligreenberg97} was used to calculate optical properties for grains with a composition of 1/3 silicates, 2/3 organic refractory materials and zero porosity, using  Mie theory, Rayleigh-Gans theory or geometric optics in the appropriate limits. The temperature in the disc now depends on the particle's  diameter, D, in addition to its distance from the star, r in AU: 
\begin{equation}
T(D,r)= \left( \frac{\langle Q_{\mathrm{abs}} \rangle _{T_*} }{\langle Q_{\mathrm{abs}} \rangle _{T(D,r)} } \right)^{1/4} T_{bb}
\end{equation}
where $\langle Q_{\mathrm{abs}} \rangle _{T_*}$ and $\langle Q_{\mathrm{abs}} \rangle _{T(D,r)}$ are the particle's absorption efficiency averaged over the stellar spectrum and the spectrum of a black body radiating at temperature, T, and $T_{bb}$ is the equilibrium temperature of the particle if it were a black body, given by:
\[
T_{\mathrm{bb}}= 278.3 \; \frac{L_*^{1/4}}{r^{1/2}}  \; \; \; \mathrm{K},
\]
where $L_*$ is the star's luminosity in solar units. The flux from the disc, at a wavelength $\lambda$, is given by:

\begin{equation}
F_{\mathrm{disc}}(\lambda)= 2.98 \times 10^{-7} \frac{P(\lambda, r) M_{\mathrm{tot}} }{\sqrt{D_{\mathrm{min}}D_c}\rho d^2} \; \; \; \mathrm{Jy},
\end{equation}
where $M_{\mathrm{tot}}$ is the mass in the disc in $M_{\oplus}$, $D_{\mathrm{min}} (\mu$m) and $D_c$ (km) are the smallest and largest particles in the disc, $\rho$ is the density of particles, in kgm$^{-3}$, d is the distance from the observer to the star, in pc,
\begin{equation}
P(\lambda, r)=\int^{D_c}_{D_{\mathrm{min}}} Q_{\mathrm{abs}}(\lambda,D) B_{\nu}(\lambda, T(D,r)) \bar \sigma(D) dD,
\end{equation}
$B_{\nu}(\lambda, T(D,r))$ is the black body flux, in JySr$^{-1}$ and $\bar \sigma (D) dD$ is the proportion of the total cross-section of the disc found in particles with sizes between $D$ and $D+dD$.

\subsection{Population of discs around A stars}

\begin{figure}
\includegraphics[width=80mm] {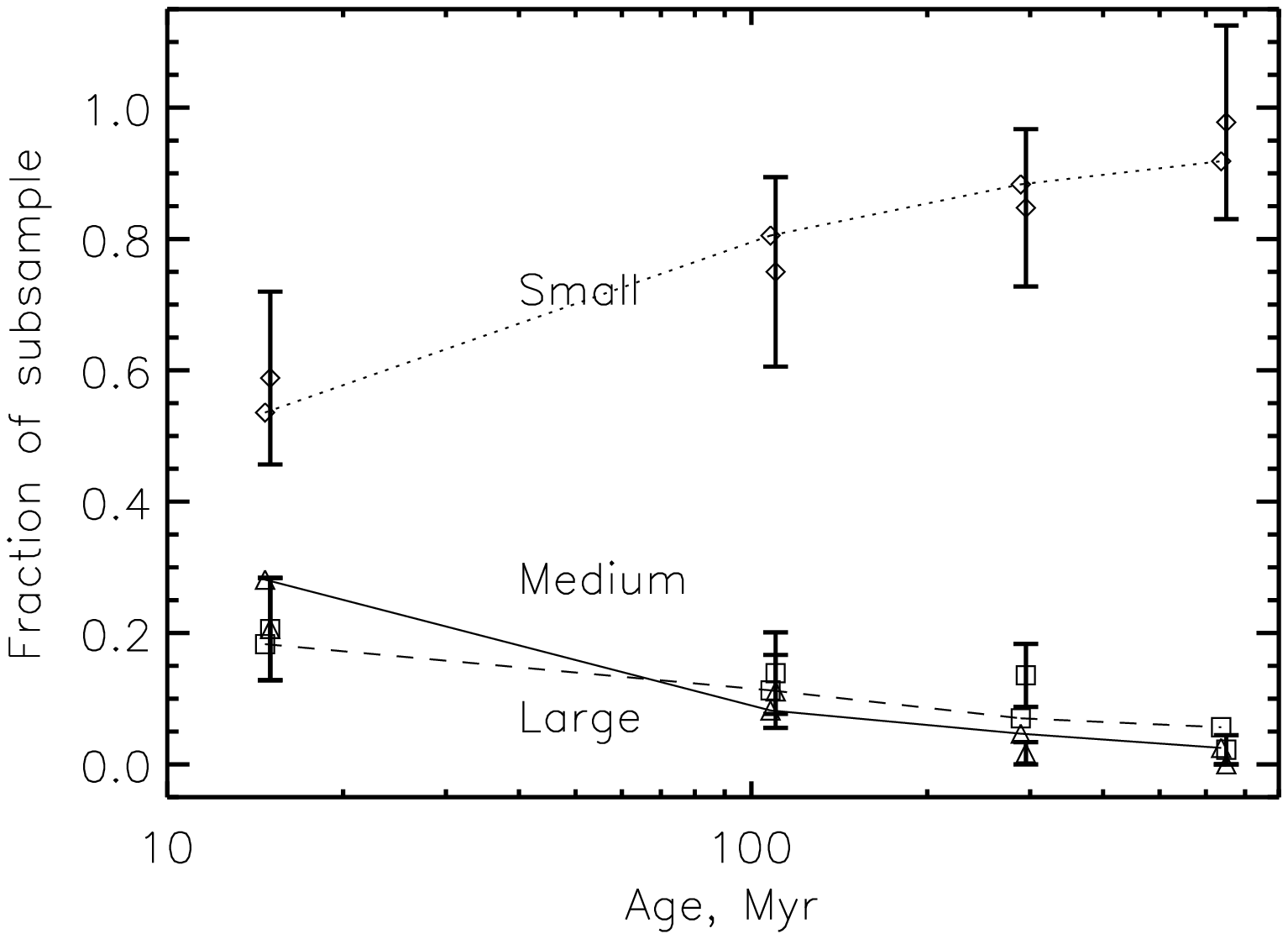}
\includegraphics[width=80mm] {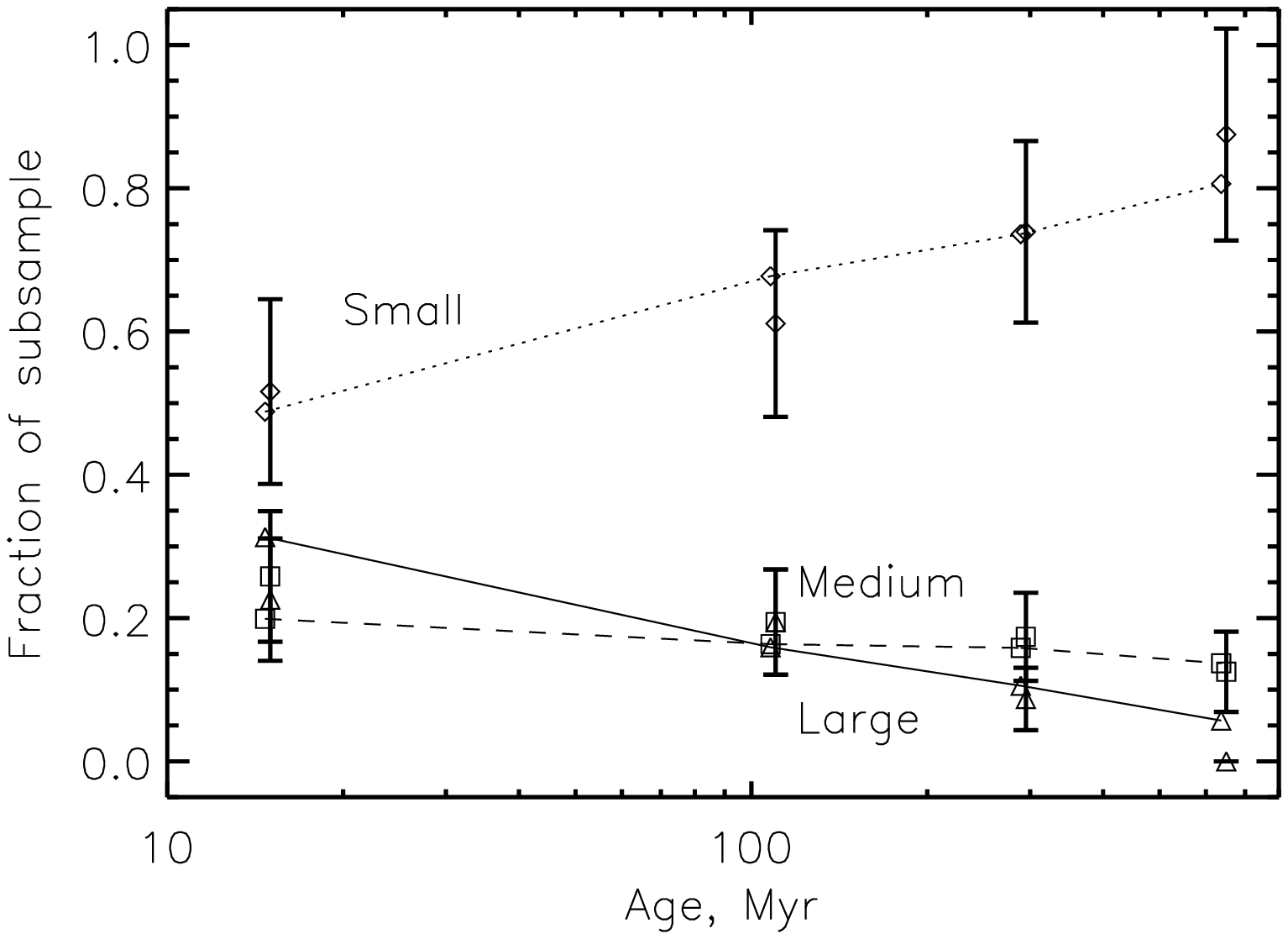}
\includegraphics[width=80mm] {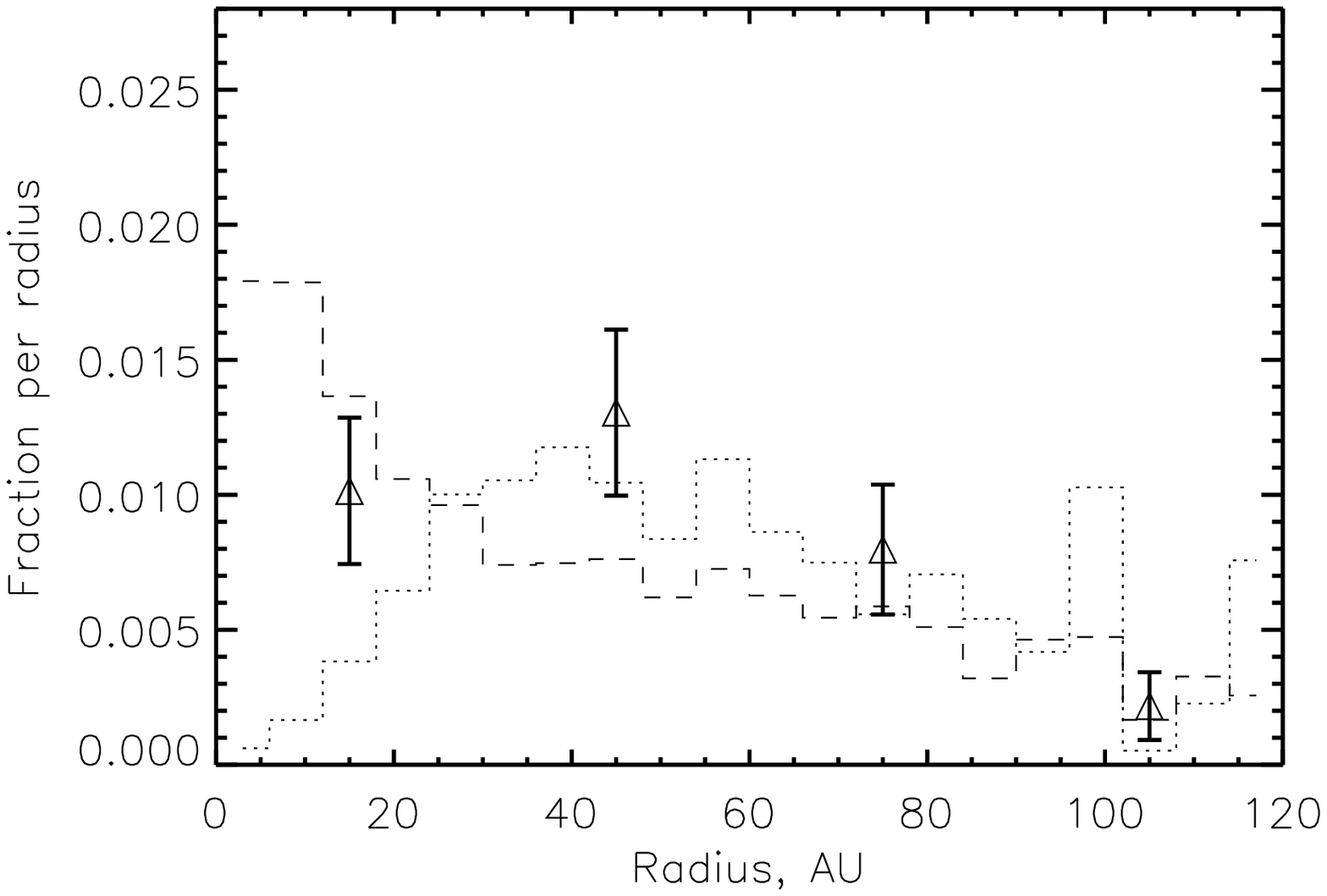}
\caption{Upper and middle panel: The fit to the observations of \citet{su06} at 24 (upper) and 70$\mu$m (middle), comparable to Fig.2 of \citet{wyatt07}. The plots show the fraction of stars with flux ratios in different age bins ($<$30Myr, 30-190Myr, 190-400Myr), at 24$\mu$m $\frac{F_{disc}}{F_*}=$ 1-1.25 (diamond: small excess), 1.25-2 (square: medium excess), $>$2 (triangle: large excess) and similarly at 70$\mu$m $\frac{F_{disc}}{F_*}$= 1-5 (diamond:small excess), 5-20 (square: medium excess), $>$20 (triangle: large excess). Observed values are shown with $\sqrt{N}$ error bars, whilst model values are joined with dotted, dash and solid lines, for small, medium and large excess. Lower: The distribution of planetesimal belt radii. The model population is shown with a dashed line, whilst the sub-sample of the model population that is detectable at both 24 and 70$\mu$m is shown with a dotted line. The observed distribution of radii are shown with triangles and $\sqrt{N}$ error bars. }
\label{fig:fit}
\end{figure}

\begin{table}
\begin{center}
\begin{tabular}{|c| c| c|}
\hline

Parameter & Orginal value & New value \\
 \hline

$M_{\mathrm{mid}}$ &10.0 &7.8 $M_{\oplus}$ \\
 $D_c$ &60.0 & 1.9km \\
 e & 0.05&0.05\\
 $Q_D^*$ &150 Jkg$^{-1}$ &150Jkg$^{-1}$\\
 $\gamma$ &0.8&0.8\\
 $r_{min}$&3AU&6AU\\
 $r_{max}$&120AU& 250AU\\
\hline
\end{tabular}
\end{center}
\end{table} 

The population of debris discs around main sequence A stars (spectral type B8 to A0) is relatively well constrained from observations by Spitzer at 24$\mu$m and 70$\mu$m \citep{rieke05, su06}. \cite{rieke05} searched a sample of 76 individual A stars, with ages between 0 and 800Myr, for excess with Spitzer at 24$\mu$m. This was extended by looking at archival IRAS data to a total of 266 stars. \cite{su06} search a sample of 160 stars for excess at both 24 and 70$\mu$m. Fig.~\ref{fig:fit} plots the stars that were found to have infrared excess in both surveys. Fig.~\ref{fig:fit} plots the fraction of stars in different age bins that are classified to have small, medium or large levels of excess. The level of excess decreases with stellar age at both wavelengths, but at a faster rate at 24$\mu$m.

\par The models in \cite{wyatt07} were fitted to these observations, using a population of 10,000 discs, with a distribution of initial masses, radii, spectral type, distance and ages. The initial masses formed a log normal distribution centered on $M_{\mathrm{mid}}$, assuming the same width as for proto-planetary discs, 1.14dex \citep{AndrewsWilliams05} and the initial radii a power law distribution, with the number of discs with radius between r and dr given by, $N(r)dr \propto r^{-\gamma}dr$, for discs between $r_{min}$ and $r_{max}$. It was assumed that the stars are randomly distributed, evenly in spectral type and age and isotropically in distance. 
In the original model the dust properties were assumed to be black-body like and the $24\mu$m and $70\mu$m statistics were fitted by adjusting the parameters, $M_{\mathrm{mid}}$, $D_c$, e, $Q_D^*$, $\gamma$, $r_{min}$ and $r_{max}$.

Here we repeated the fit, using the updated formulation that incorporates the optical properties of realistic grains, to find new values for this parameter set. As discussed in \cite{wyatt07} a degeneracy in the model means that these parameters are not uniquely constrained. The fit to the data is good so long as both $D_c^{1/2} Q_D^{*,5/6} e^{-5/3}$ (Eq. 15 of \cite{wyatt07}) and $M_{tot,mid} D_c^{-1/2}$ (Eq.16 of \cite{wyatt07}) remain constant. The former can be explained by the example of a massive disc composed of stronger bodies that evolves on the same timescale as a less massive disc of weaker bodies. Alternatively the later can be explained by the fact that a massive disc that contains more of its mass in larger bodies (i.e. a larger value for $D_C$) has the same dust mass and therefore is equivalently bright as a less massive disc where $D_C$ is smaller. Here we chose to keep $Q_D^*$ and e unchanged at 150Jkg$^{-1}$ and 0.05 respectively, without any loss of generality and performed a fit to $M_{mid}$, $D_C$, $\gamma$, $r_{min}$ and $r_{max}$. 

In order to fit the observations with a new population calculated using the emission properties of realistic grains, we want every disc to evolve in the same way as every disc in the old population, calculated using a black body approximation. The optical properties of realistic grains mean that they are hotter and emit less efficiently at longer wavelengths. Therefore in order for the flux ratio $\frac{F_{\mathrm{disc}, 24\mu m}}{F_{\mathrm{disc}, 70\mu m}}$ to remain the same, disc radii must increase. In fact a good fit is achieved by adjusting the disc radii from the values in the original model, r2470, to a new value, $r=X_{2470}r_{2470}$, so long as the other parameters are also adjusted accordingly; thus we keep $\gamma$ at 0.8, whilst $r_{min}$ and $r_{max}$ increase by a factor $X_{2470}$. Then, in order to keep the flux from each disc constant, its fractional luminosity (f), the ratio of the luminosity of the disc to the luminosity of the star, should remain unaltered. Grains are larger than the emission wavelength of starlight on the main sequence such that they absorb starlight efficiently and $f  \propto \frac{ M_{tot}}{r^2 D_c^{1/2}D_{min}^{1/2}}$. Therefore we need to adjust $M_{tot}$ and $D_C$ keeping $\frac{M_{tot}}{\sqrt{D_c}r^2}$ constant. The maximum fractional luminosity that a disc of a given age can have due to its collisional evolution should also remain constant and thus using Eq.~\ref{eq:mmax} $r^{7/3}D_c^{1/2}$ is also a constant. Together these mean that $D_c$ and $M_{mid}$ are altered by $X_{2470}^{-14/3}$ and $X_{2470}^{-1/3}$ respectively.

\begin{figure}
\includegraphics[width=80mm] {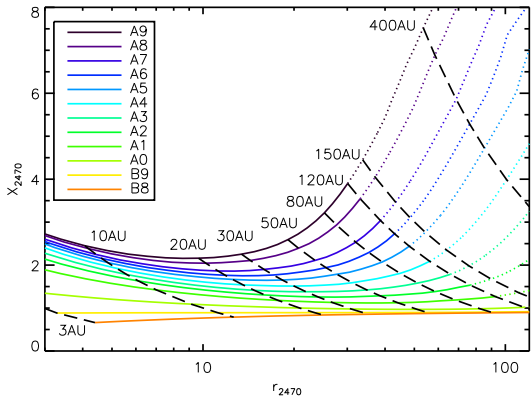}
\caption{The ratio of the radius (r) of a disc calculated using realistic grains to the radius inferred from the flux at $24\mu$m and 70$\mu$m, assuming black body emission (r$_{2470}$), as a function of r$_{2470}$.  }
\label{fig:ratio}
\end{figure}

 The conversion between r and r$_{2470}$ was determined for realistic grains with the prescribed size distribution around main sequence stars and is shown in Fig.~\ref{fig:ratio}. There is a functional dependence of $X_{2470}$ on r$_{2470}$ and spectral type that can be readily understood.
For a given spectral type, $X_{2470}$ has a minimum at intermediate radii 
 but increases at small and large radii. The latter arises because the cooler temperatures at larger radii mean that the emission at these wavelengths is on the Wien side of the black body spectrum. This means that the small increase in the temperature of the grains causes a larger increase in the flux ratio at these wavelengths and hence the radius inferred, r$_{2470}$. This is compounded by the fact that the temperature of blow-out grains falls off more slowly with radius than that of black body grains (e.g.~\cite{hr8799su}). The increase at small radii arises because the emission efficiency of realistic grains falls off with wavelength, such that the emitted spectrum appears steeper (i.e. hotter) than the true grain temperatures. This effect is more important where the spectrum is steeper, i.e. where discs are hotter. All of these effects are more pronounced for later spectral type stars because the blow-out size is smaller (Eq.~\ref{eq:dbl}) and therefore there is a larger population of grains whose properties depart from black body.

 $X_{2470}$ was found self-consistently using Fig.~\ref{fig:ratio} to give an average value of 2.1. The above discussion suggests that an equally good fit to the statistics could be obtained with $M_{\mathrm{mid}}=7.8 M_{\oplus}$ and $D_c=1.9$km. Indeed, as is shown in Fig.~\ref{fig:fit} this is found to be the case, with a total $\chi^2_{24,70,r}$ of 16.0 compared to 9.8 of \cite{wyatt07}.

\section{Models of debris discs around post-main sequence A stars}
\label{sec:models}

The steady state models for the evolution of the disc due to collisions described in \S \ref{sec:modelmw} are extended to include the effects of the evolution of the star, described in \S \ref{sec:star}. The evolution of individual discs and changes to their properties, are discussed. Individual discs follow different evolutionary paths depending on their properties. The plots presented in this section are representative, and look at the evolution of a disc around a 2.9$M_{\odot}$ (equivalent to A0) star, with solar metallicity ($Z=0.02$), initial masses in the disc of 1.0, 10.0 and 100.0 $M_{\oplus}$ and initial radii of 10, 50 and 100 AU, although the population models discussed in Sec.~\ref{sec:pop} use the evolution for the appropriate disc and stellar parameters.

\subsection{Radius evolution}

As the star loses mass, the semi-major axes of orbiting bodies will increase. 
The star loses the majority of its mass in $10^5$yr at the tip of the AGB. For the majority of discs this mass loss timescale is much longer than the orbital timescales of particles and therefore the mass loss can be considered to be adiabatic. Only for the discs at much larger radii do the orbital timescales start to approach the mass loss timescales and the assumption of adiabatic mass loss may break down, inducing eccentric orbits, for example a body initially in a circular orbit at 100AU will gain an eccentricity of 0.1 if mass loss rates reach $10^{-4}M_{\odot}yr^{-1}$.

 However these effects are small and are ignored in the current models, where the evolution of the radius (r) due to adiabatic mass loss is given by \cite{villaverlivio} :
\begin{equation}
\label{erad}
r(t)=\frac{r(0) M_*(0)}{M_*(t)}
\end{equation}
Essentially the radius of the disc switches from an initial to a final value during the $\sim 10^5$ yr of extreme mass loss, at the tip of the AGB, as shown in the upper panel of Fig.~\ref{mtot}.

\subsection{ Mass in the disc and the collisional lifetime}
\label{sec:mtot}
\begin{figure}
\includegraphics[width=80mm] {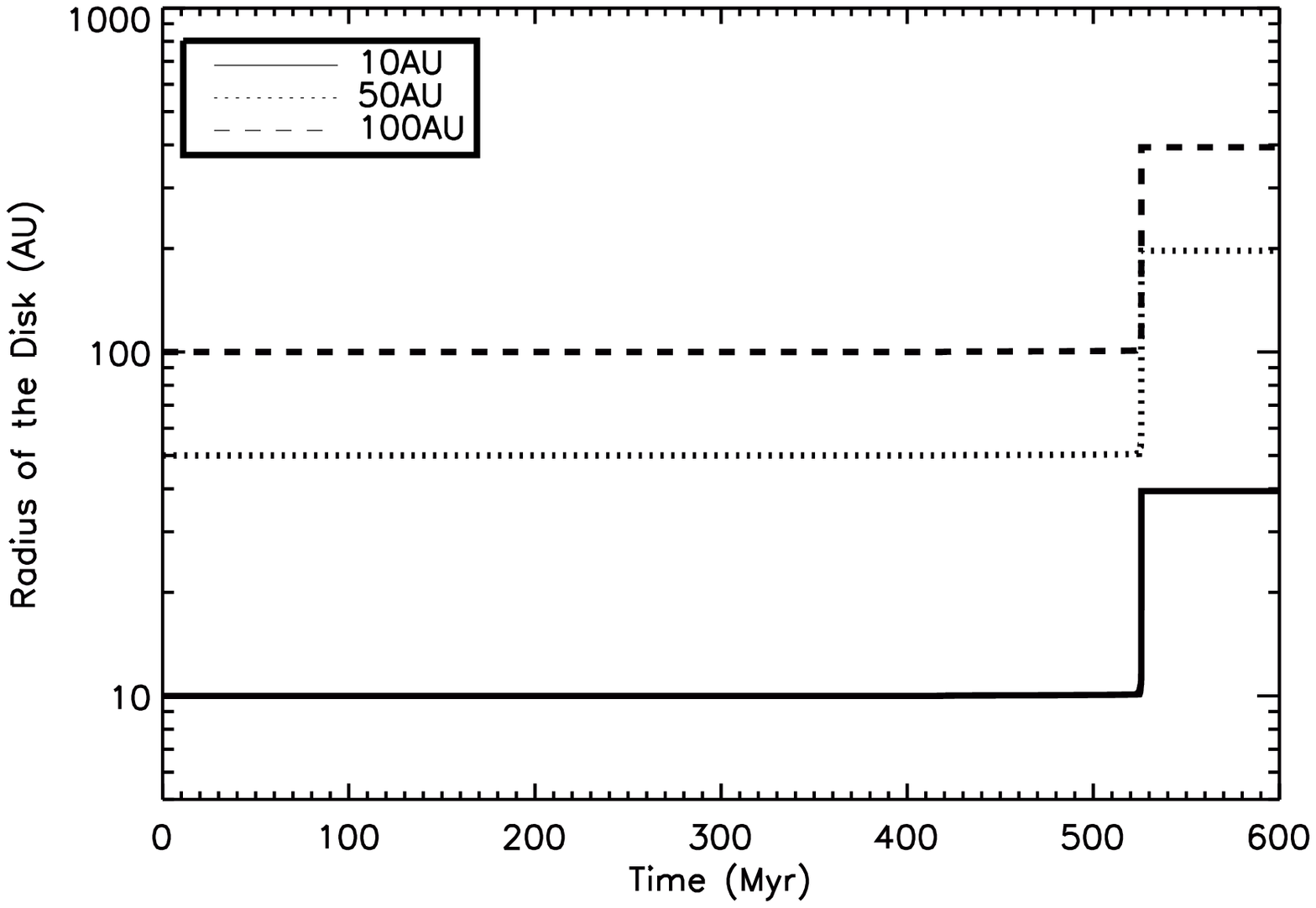}
\label{frad}
\includegraphics[width=80mm] {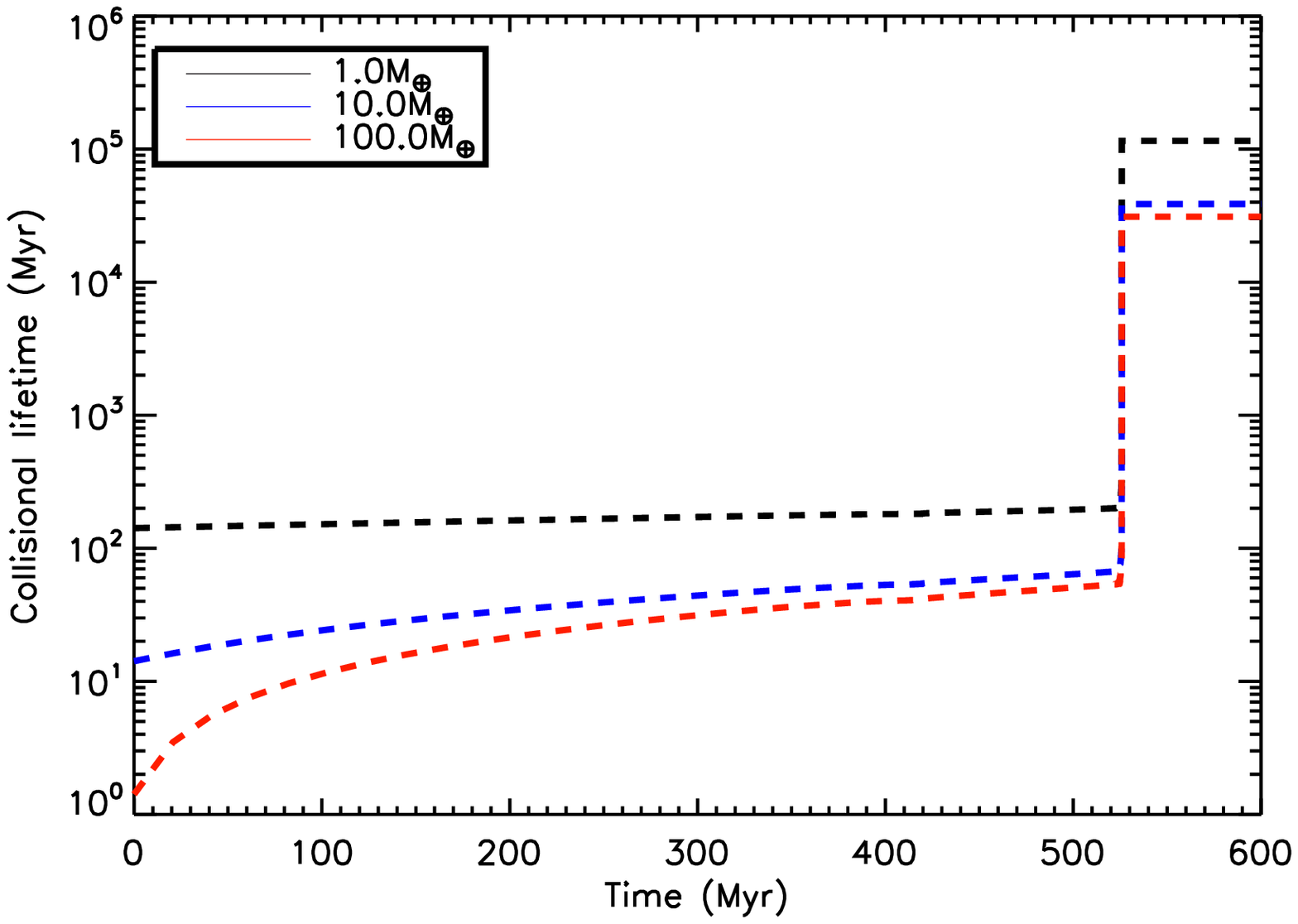}
\label{fig:tc}
\includegraphics[width=80mm] {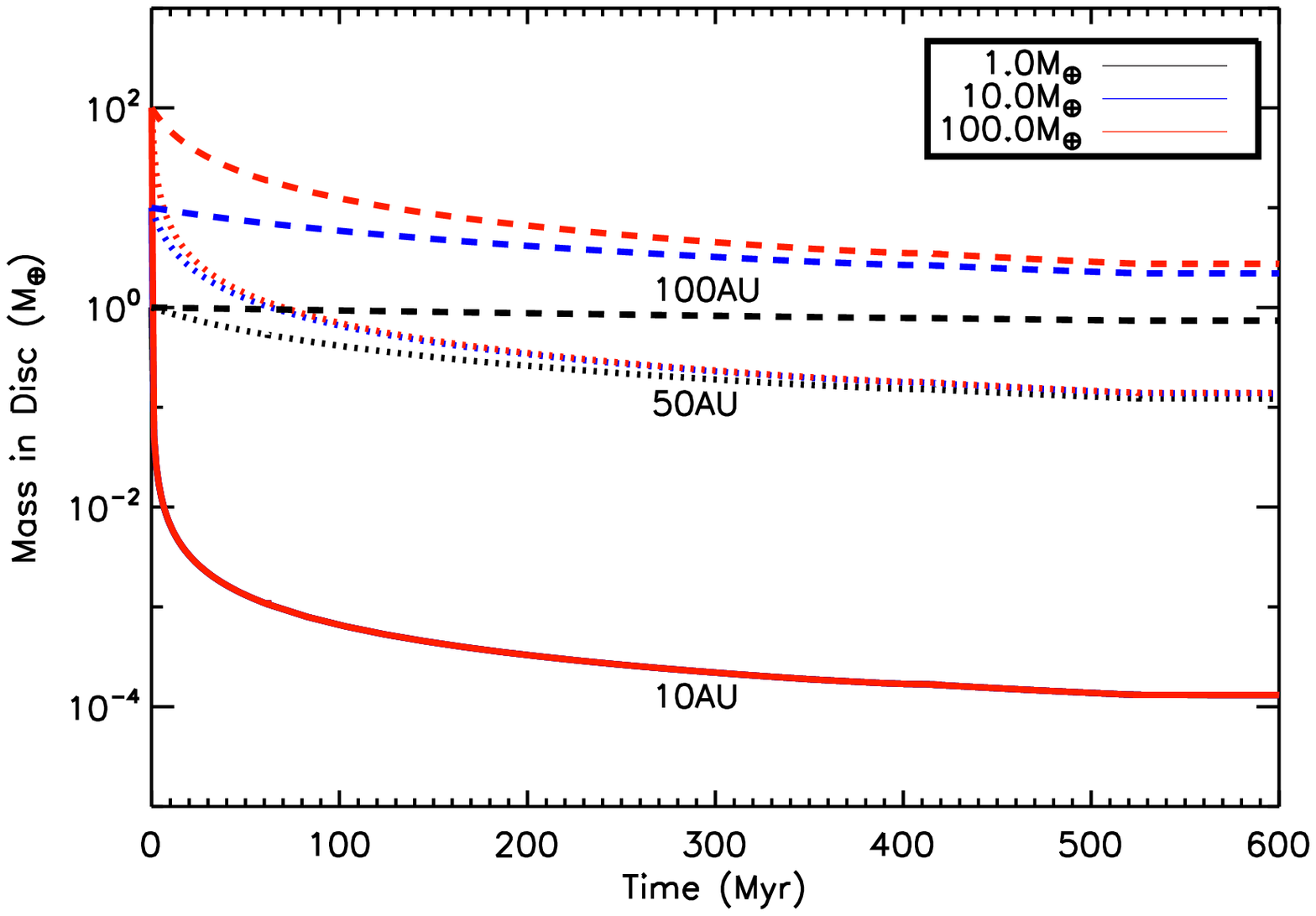}
\label{fsfd}
\caption{The evolution of the disc radius (upper), collisional lifetime (middle) and total disc mass (lower), for different initial disc masses, $1.0M_{\oplus}$ (black), $10.0M_{\oplus}$ (blue), $100.0 M_{\oplus}$ (red), and radii, 100AU (dashed line), (bottom plot only 50AU (dotted line), 10AU (thick line)), around a 2.9$M_{\odot}$ star. }
\label{mtot}
\end{figure}

As discussed in \S \ref{sec:modelmw}, the mass in the disc is dominated by the largest particles, and hence the timescale on which the mass is depleted is dependent on the collisional lifetime of the most massive particles, $t_c$ (Eq.~\ref{eq:tc}). The evolution of $t_c$ is shown in the middle panel of Fig.~\ref{mtot}. Collisions occur most frequently ($t_c$ is shorter) in the most massive discs, closest to the star. The collisional lifetime increases significantly when the star loses mass, and is given by:
\begin{equation}
{
t_c=\frac{t_c(0) M_{tot}(0) M_*^{17/3}(0)}{M_*(t)^{17/3}M_{tot}(t)}.}
\end{equation}
  Once the mass in the star changes as a function of time, the evolution of the mass in the disc is no longer given simply by Eq.~\ref{eq:dmtot}, instead:
\begin{equation}
\frac{dM_{\mathrm{tot}}}{dt}=\frac{-M_{\mathrm{tot}}}{t_c} \propto \frac{M^2_{\mathrm{tot}}(t) M_*^{4/3}(t)}{r^{13/3}(t)}
\end{equation}
Using the expression for r(t), 
\begin{equation}
\int\frac{dM_{\mathrm{tot}}}{M_{\mathrm{tot}}^2} \propto \int{ M_*^{17/3}dt}
\end{equation}
Therefore 

\begin{equation}
\label{mtoteq}
M_{\mathrm{tot}}=\frac{M_{\mathrm{tot}}(0)}{1+M_{\mathrm{tot}}(0) K \int M_* ^{17/3} dt }
\end{equation}

where \begin{equation}
K=[1.4\times10^{-9}(\frac{dr}{r})e^{-5/3}D_c Q_D^{* 5/6}r(0)^{13/3}M_*(0)^{13/3}]^{-1}.
\end{equation}

However the mass in the star is approximately constant until the AGB, and up until this point Eq.~\ref{eq:dmtot} is valid. In fact up to the end of the AGB, the evolution of the mass in the disc is similar to that on the main sequence in that $M_{tot}$ remains approximately constant for discs with longer collisional lifetimes (larger radii) and tends to $M_{max}$ (Eq.~\ref{eq:mmax}) for discs with shorter collisional lifetimes (small radii), as shown in the bottom panel of Fig.~\ref{mtot}.  Once the star loses mass on the AGB, the collisional lifetime of larger objects increases so much that it approaches the Hubble time, even for close-in discs (see the middle panel of Fig.~\ref{mtot}).  Collisional evolution is no longer significant for the total disc mass, which remains constant throughout the white dwarf phase.

\begin{figure}
\subfigure{\includegraphics[width=80mm] {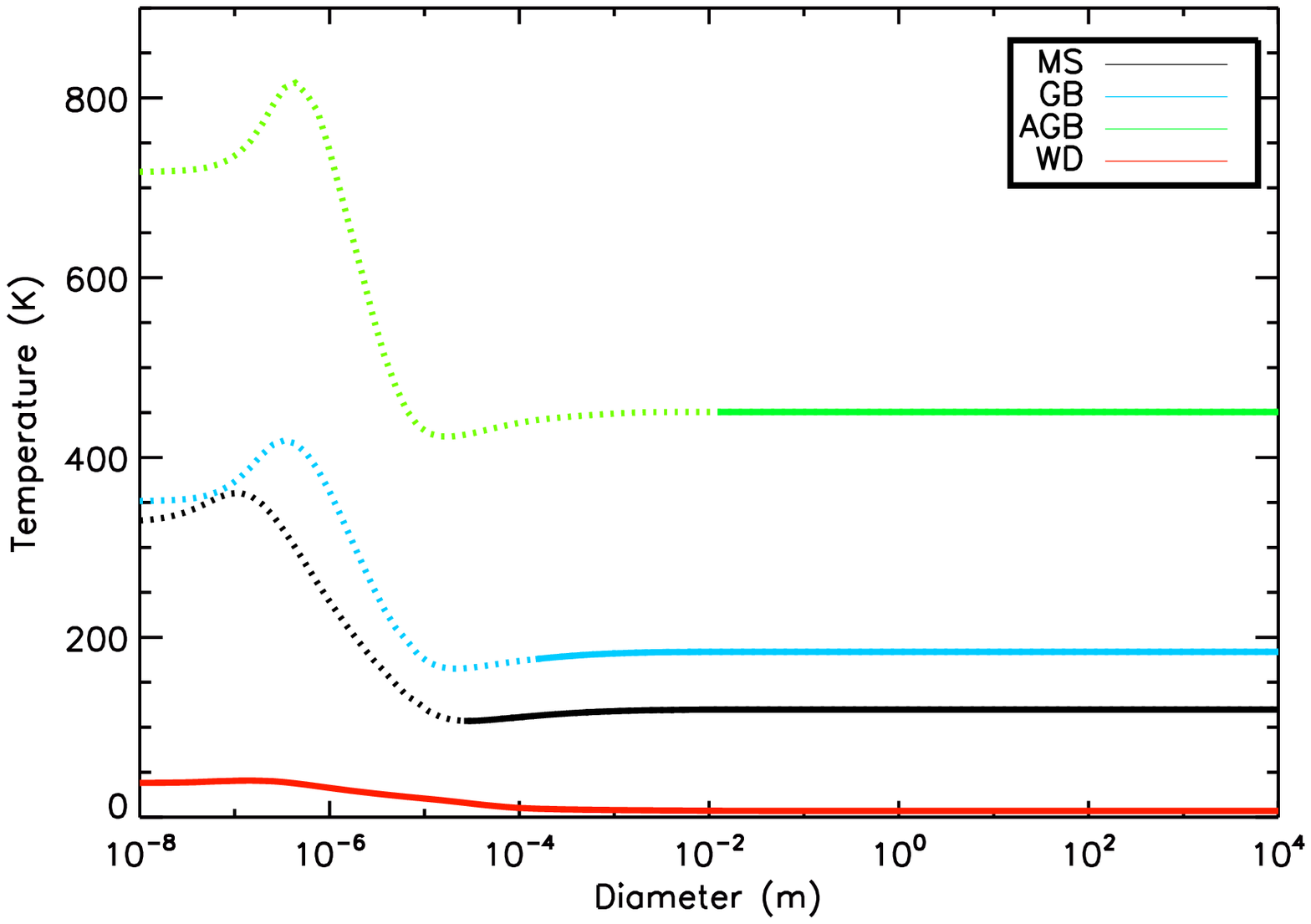}}

\subfigure{\includegraphics[width=80mm] {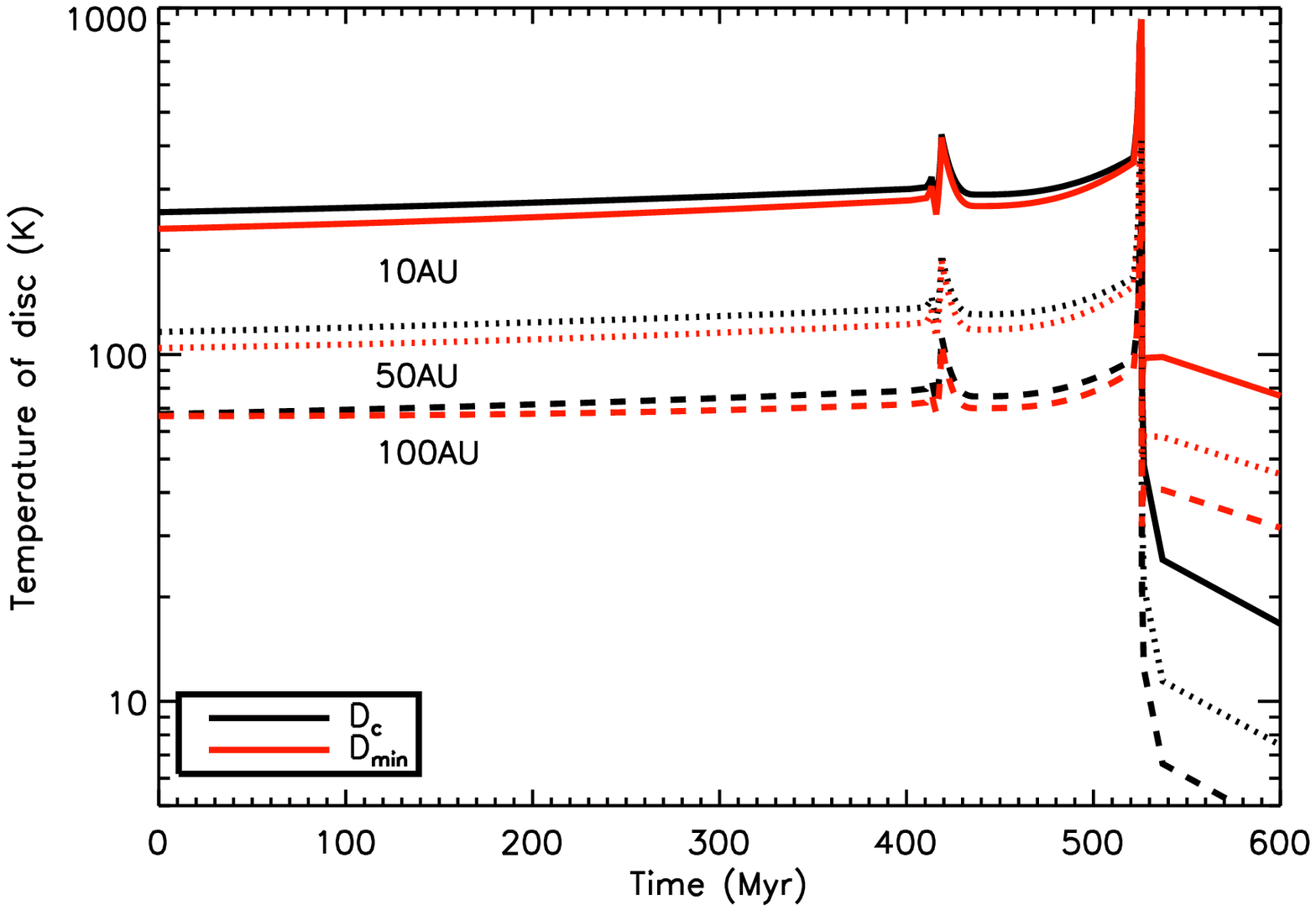}}

\caption{Upper: The temperature of particles in the disc as a function of particle diameter, for a disc initially at 50AU around a $2.9M_{\odot}$ star, around a main sequence star ($L_*= 190 L_{\odot}$), giant star ($L_*= 500 L_{\odot}$), a horizontal branch star ($L_*= 150 L_{\odot}$), an AGB star ($L_*= 1.5 \times 10^{4} L_{\odot}$) or white dwarf ($L_*= 7\times 10^{-3} L_{\odot}$). The solid lines show bound grains that are included in the model, whereas the dotted lines are unbound grains that are excluded from the model. 
Lower: The evolution of the temperature of black body grains ($D_c$) and the smallest grains in the disc ($D_{min}$), for discs initially at 10AU (thick line), 50AU (dotted line), 100AU (dashed line) around a $2.9M_{\odot}$ star. }

\label{fig:temp}
\end{figure}

\subsection{Temperature of the disc}
\label{sec:temp}

Particles in the disc are heated by stellar radiation. Their temperature is a balance between absorption and emission and is a function of particle diameter, as shown in the upper panel of Fig.~\ref{fig:temp}. Large grains emit efficiently and act like black bodies, whilst medium sized grains ($\sim \mu$m) emit inefficiently and are therefore hotter than black body. The smallest ($<\mu$m) grains emit and absorb inefficiently and reach a temperature that is independent of grain size. 

To illustrate the evolution of the temperature of particles in the disc we show in the lower panel of Fig.~\ref{fig:temp} the change in the temperature of black body particles (appropriate for large grains) and the smallest grains in the disc of size $D_{min}$ (as calculated in \S \ref{sec:small}) as the star evolves. This follows the luminosity of the star. Along the main sequence the temperature of the disc is relatively constant, but it increases up to several hundred Kelvin (depending on disc radius) 

 as the star's luminosity increases on the RGB and AGB.  The temperature of the disc falls dramatically as the star becomes a white dwarf, mostly because the stellar luminosity drops by several orders of magnitude, but also because the discs are now further from the star. 
For this example mass star, the only epoch when the temperature of any particle in the disc greatly exceeds the black body temperature is for white dwarfs, however for later spectral type (lower mass) stars the smaller grains may be hotter than blackbody through all phases of stellar evolution.

\begin{figure}
\includegraphics[width=80mm] {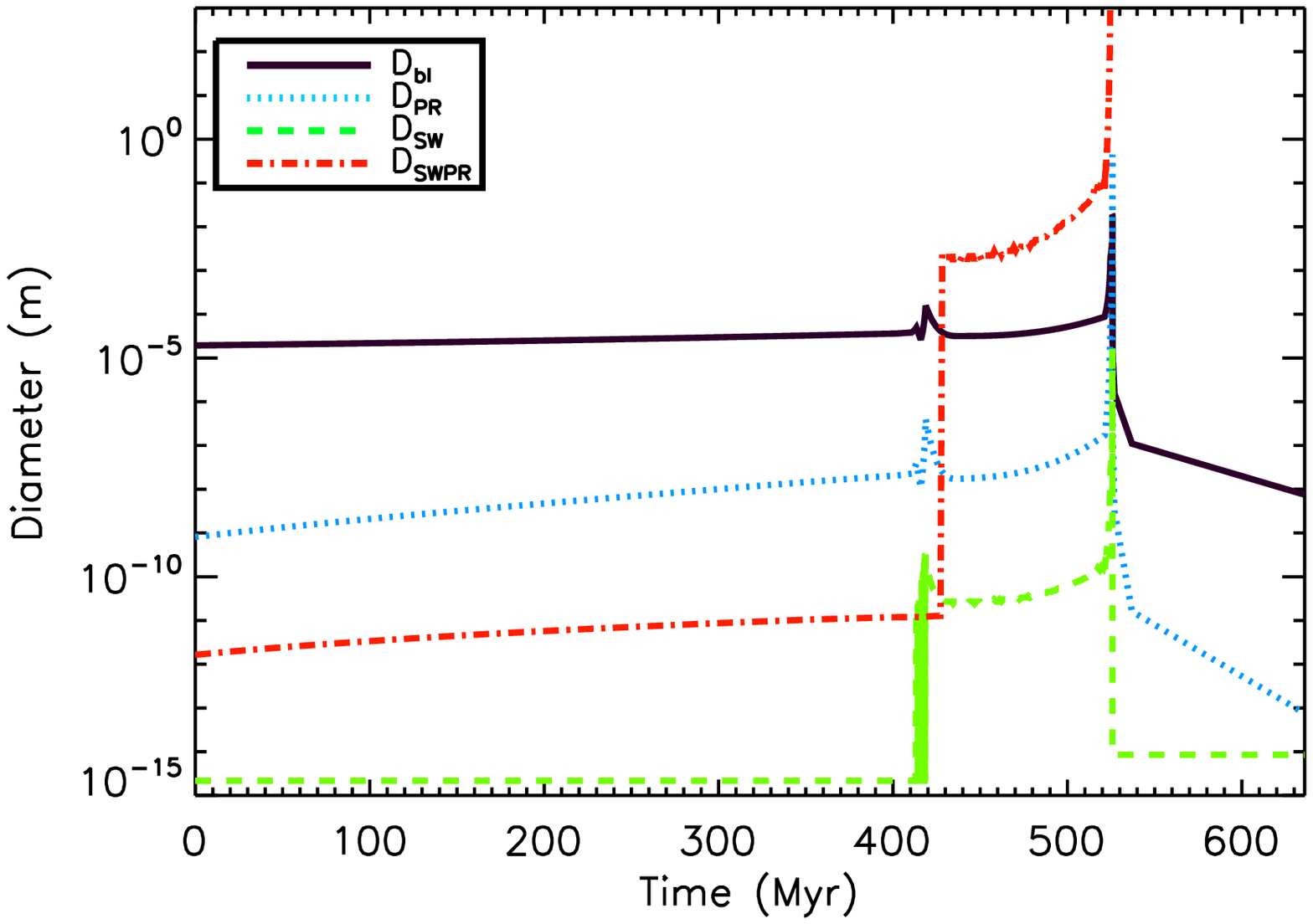}
\includegraphics[width=80mm] {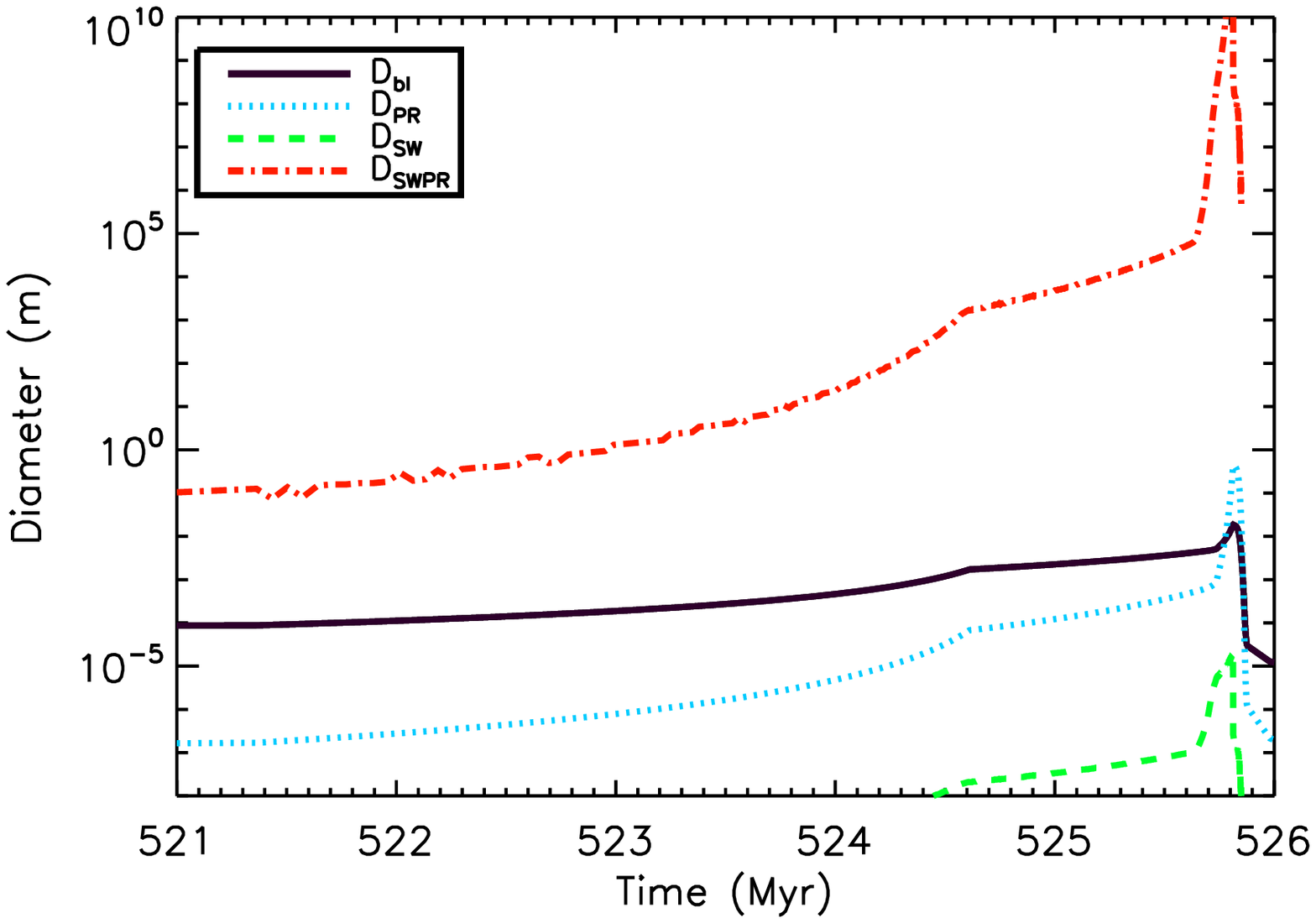}
\caption{The limiting diameter below which objects are removed by either radiation pressure ($D_{bl}$), PR-drag ($D_{PR}$), stellar wind pressure ($D_{SW}$) or stellar wind drag ($D_{SWPR}$), calculated using Eq.~\ref{eq:dbl}, Eq.~\ref{eq:dpr}, Eq.~\ref{eq:dsw} and Eq.~\ref{eq:dswpr}, in a disc initially at 100AU with a mass of $10M_{\oplus}$,around an evolving 2.9$M_{\odot}$ star. The lower panel shows a zoom-in on the AGB. $D_{SWPR}$ overestimates the diameter bodies that are removed by stellar wind drag as it does not take into account the finite AGB lifetime.}
\label{fig:diam_lim}
\end{figure}

\subsection{Smallest particles in the disc}
\label{sec:small}
The size of the smallest particles ($D_{min}$) in the collisional cascade was determined in the \cite{wyatt07} models and in \S \ref{sec:modelmw} by radiation pressure (the blow-out size Eq.~\ref{eq:dbl}). This is correct for most discs around main sequence stars, however there are several other forces that can remove small particles from the disc, including Poynting-Robertson drag, stellar wind pressure, stellar wind drag and sublimation. In the models presented here, $D_{min}$ is determined by whichever of these removes the largest diameter objects at a given epoch. In the following section all five processes are discussed and compared such that $D_{min}$ can be determined for every disc during its evolution. The outcome is summarised in Fig.~\ref{fig:diam_lim}.

\subsubsection{Radiation pressure}
\label{sec:radp}
\begin{figure}
\includegraphics[width=80mm] {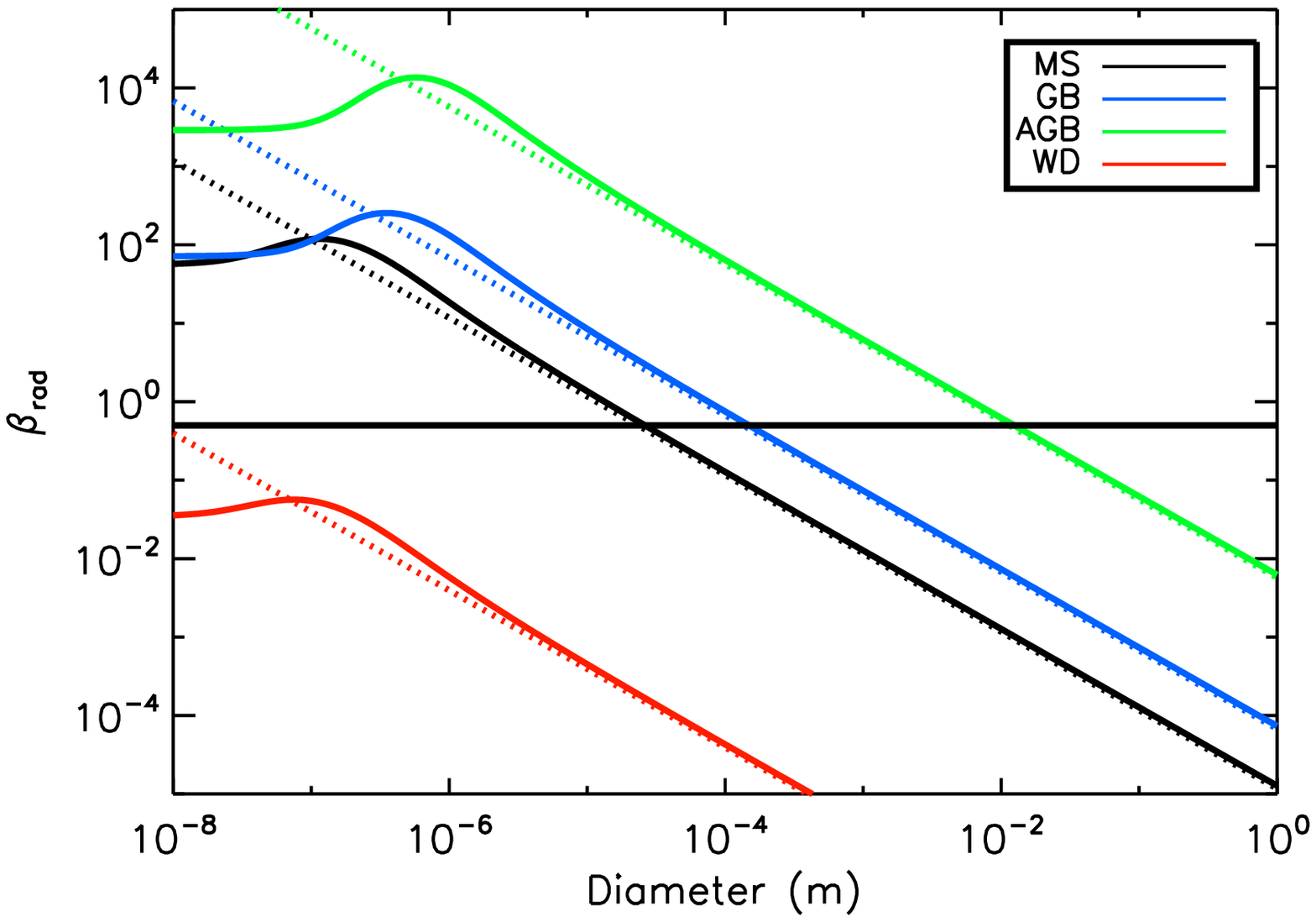}
\includegraphics[width=80mm] {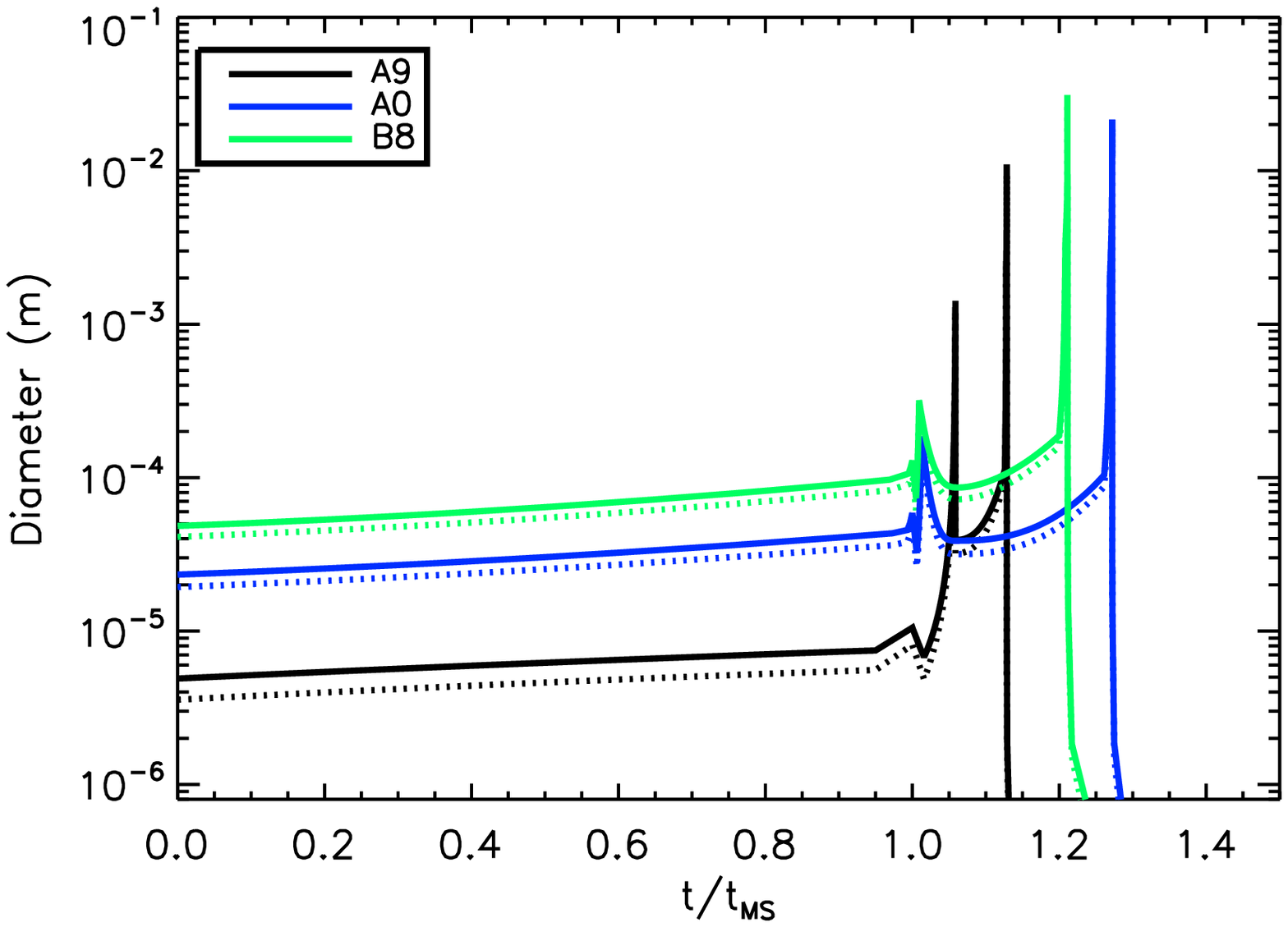}
\includegraphics[width=80mm] {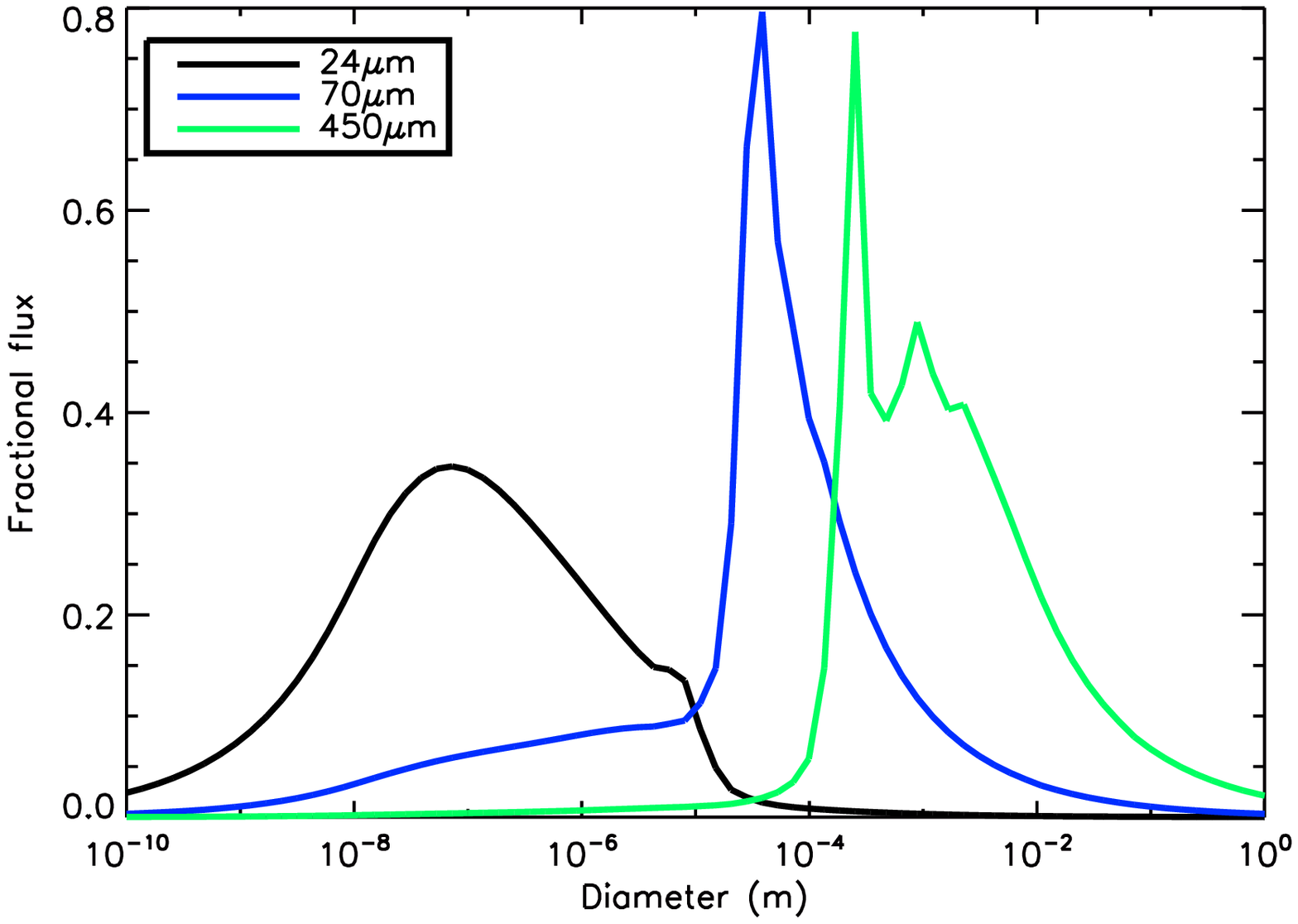}
\caption{
The effect of radiation pressure on the disc. Calculations with $\langle Q_{\mathrm{pr}}\rangle =1$ are shown with dotted lines, whilst the solid lines show a more realistic calculation. 
Upper: The ratio of the radiational to gravitational forces ($\beta_{rad}$) for different diameter particles in a disc around a main sequence star ($L_*= 190 L_{\odot}$), giant star ($L_*= 500 L_{\odot}$), AGB star ($L_*= 1.5 \times 10^{4} L_{\odot}$) or white dwarf ($L_*= 7\times 10^{-3} L_{\odot}$). The horizontal black line shows $\beta_{rad}=0.5$. Particles with $\beta_{rad}>0.5$ are removed from the system by radiation pressure. The maximum value of $\beta_{rad}$ is less than 0.5 around a white dwarf, once $L_{WD}$ falls below $\sim 0.15 L_{\odot}$.
Middle: The change in the blow-out diameter of realistic grains around 1.67, 2.9 and 3.8 M$_{\odot}$ or A9, A0 and B8 stars. 
For both plots the dotted lines show calculations with $\langle Q_{\mathrm{pr}}\rangle =1$, which do not vary significantly from the solid lines, which include a more realistic calculation of $\langle Q_{\mathrm{pr}}\rangle $, apart from for small diameter particles.  
Lower: The fraction of the flux per unit log diameter, defined such that the area under the curve is 1, for a disc at 100AU, around an evolved 2.9$M_{\odot}$ star as a white dwarf with a cooling age of 1Myr. }
\label{fig:beta}

\end{figure}

 Radiation pressure is a radial force which acts in the opposite direction to the star's gravity. The ratio of radiation forces to the gravitational forces is given by:
\begin{equation}
\beta_{\mathrm{rad}}=\frac{F_{\mathrm{rad}}}{F_{\mathrm{grav}}}= 0.43 \frac{\langle Q_{\mathrm{pr}}\rangle L_*}{M_* D},
\label{eq:beta}
\end{equation}

where $\langle Q_{\mathrm{pr}}\rangle$ is the radiation pressure efficiency for grains of a given diameter D, in $\mu$m, averaged over the stellar spectrum. The dependence of $\beta_{rad}$ on particle diameter, for realistic grains, is shown in the upper panel of Fig.~\ref{fig:beta} at different epochs. The peak in $\beta_{rad}$ occurs at a size comparable to the peak wavelength in the stellar spectrum. The dotted line shows the approximation to $\beta_{rad}$ used in this work for which $\langle Q_{\mathrm{pr}}\rangle=1$. It only deviates from the more realistic calculation at the smallest particle sizes and since these small particles are generally removed from the disc, apart from during the white dwarf phase, this is considered a reasonable approximation.

If $\beta_{\mathrm{rad}}>1$ the radiation forces are greater than the gravitational forces and particles are unbound. However for small particles produced in collisions radiation pressure causes their orbits to differ from that of the parent body, resulting in a hyperbolic orbit if $\beta_{\mathrm{rad}}>0.5$ \citep{burns}. Radiation pressure also causes bound grains with $0.1<\beta_{\mathrm{rad}}<0.5$ to have eccentricities greater than their parents. This causes a perturbation to the size distribution from that assumed in Eq.~\ref{eq:sizedistrib} (e.g. \cite{strubbechiang, thebault07}), but the resulting disc flux can be approximated by assuming the eccentricity to be constant for all particles and ignoring the effect on the size distribution, which is assumed to be truncated at $D_{bl}$ calculated using Eq.~\ref{eq:dbl}.

\par
The middle panel in Fig.~\ref{fig:beta} shows how the blow-out size ($D_{bl}$) changes as the star evolves for stars with different initial masses. The blow-out size follows the luminosity evolution of the star, increasing both on the RGB and AGB, up to $\sim$cm in size, before decreasing significantly as the star becomes a white dwarf. The stellar luminosity of a white dwarf is so low that once the star has cooled sufficiently ($L_{WD}\le 0.1L_{\odot}$), $\beta_{rad}$ is never above 0.5 (as seen in Fig.~\ref{fig:beta}) and no grains are removed from the system by radiation pressure. The change in stellar luminosity or $\beta_{rad}$ with time causes particles to spiral inwards, changing its semi-major axis by:
\begin{equation}
a(t)=\frac{a(0)(1-\beta_{rad}(0))}{(1-\beta_{rad}(t))},
\end{equation}
 where a(0) and $\beta_{rad}(0)$ are the semi-major axis and ratio of the radiation pressure to the stellar gravity at the start of the white dwarf phase. No particles ever reach the star by this process, the maximum fractional change in semi-major axis is $(1-\beta_{rad}(0))$.

\par  This leaves the question of what, if anything, removes the smallest particles from collisional cascades in discs around white dwarfs, a problem which also exists for M-dwarfs \citep{Mdwarfs09}. It is possible that magnetic effects or interactions with the interstellar medium remove the smallest particles in discs around white dwarfs. In this work, however, the fate of the smallest particles is left as an open question. The bottom panel of Fig.~\ref{fig:beta} shows that, should they exist, particles smaller than $D_{min}=10^{-8}$m contribute negligibly to the total flux in the wavebands considered here. This is because, despite the size distribution of Eq.~\ref{eq:sizedistrib} meaning that such grains contain the majority of the cross-sectional area in the disc, such small grains also have extremely low emission efficiencies at longer wavelengths. For example even for the extreme case of a disc at 100AU, around an evolved 2.9$M_{\odot}$, with a white dwarf cooling age of 1Myr and a size distribution that extends down to 10$^{-10}$m, the contribution of particles less than $10^{-8}$m in size to the $24 \mu$m disc flux is only $20\%$. Thus for practical purposes we set $D_{min}=10^{-8}$m.

\subsubsection{Poynting-Robertson Drag}
\par
Radiation forces also oppose the velocity of an orbiting dust particle, reducing its angular momentum and causing it to spiral inwards, by Poynting-Robertson drag (PR-drag), changing its radius by order itself on timescales of
\begin{equation}
t_{\mathrm{pr}}= 1.4 \times 10^{-6} \frac{r^2 \rho D }{L_* \langle Q_{\mathrm{pr}}\rangle} \; \; \; \mathrm{Myr}.
\label{eq:tpr}
\end{equation}

 Poynting-Robertson drag is only relevant for particle sizes for which the PR-drag timescale is significantly shorter than the collisional lifetime, since otherwise the particles are destroyed by collisions before they have had the opportunity to migrate. 
 Assuming that the size distribution extends down in size indefinitely according to Eq.~\ref{eq:sizedistrib} and that PR-drag lifetime varies according to Eq.~\ref{eq:tpr}, both of which are valid in the regime where radiation pressure is negligible, it is possible to derive a condition for the diameter, $D_{PR}$, at which the collisional cascade is truncated by PR-drag, by comparing the collisional lifetime of the smallest grains to their PR-drag lifetime:
\begin{equation}
D_{PR}=8.63 \times 10^{-23} \frac{L_*^2(\frac{dr}{r})^2 r^{7/3} Q_D^{* 5/3}D_c } { M_*^{8/3} M_{tot}^2 e^{10/3}}   \;\;\; \mu m. 
\label{eq:dpr}
\end{equation}

In Fig.~\ref{fig:diam_lim} $D_{PR}$ is compared to $D_{bl}$ (Eq.~\ref{eq:dbl}). $D_{PR}$ is always smaller than $D_{bl}$ for the disc initially at 100AU with 10$M_{\oplus}$ shown, such that collisions and radiation pressure dominate over PR-drag which can therefore be ignored, as was previously shown in \cite{wyatt05}. A similar analysis for discs of different mass and radii around different mass stars shows that PR-drag can always be ignored except for close-in discs or those low in mass.

PR-drag in discs around white dwarfs is of particular interest as a possible mechanism to remove the smallest grains. Of potential importance is the fact that objects larger than $D> \frac{22.4mm}{r(0)^2}$ can never reach the star due to PR-drag. This is because the luminosity of the white dwarf decreases, and thus the rate at which objects spiral in decreases with time. The critical size is calculated by solving for the rate of change of semi-major axis, a, due to PR-drag, for a zero eccentricity particle, given by \citep{burns}:
\begin{equation}
\langle\frac{da}{dt}\rangle=-\frac{3L_{wd}}{8 \pi \rho c^2 D a}
\end{equation}
where $L_{wd}$ is given by Eq.~\ref{LWD}. However since collisions still occur on faster timescales than PR-drag can act, even for the smallest particles present ($10^{-8}$m, as is seen in the upper panel of Fig.~\ref{fig:diam_lim}), it is not expected that PR-drag has a significant effect on debris discs around white dwarfs. 

\begin{figure}
\includegraphics[width=80mm] {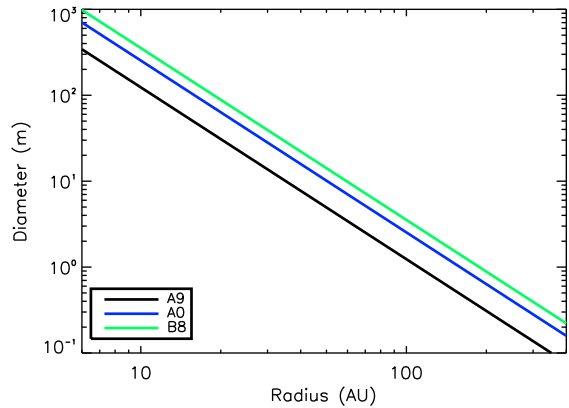}
\caption{The maximum diameter object that can be removed by stellar wind drag, throughout the disc's evolution, ignoring collisions, as a function of disc radius, for a disc around a 1.67$M_{\odot}$ (A9), 2.9$M_{\odot}$ (A0) or 3.8$M_{\odot}$ (B8) star.}
\label{fig:maxdiamsw}
\end{figure}

\begin{figure}

\includegraphics[width=80mm] {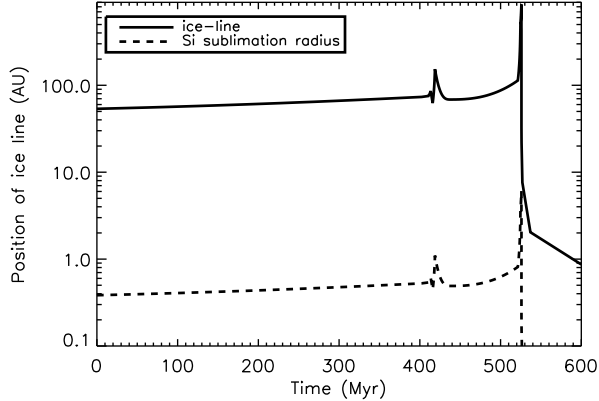}
\caption{The change in position of the ice-line and silicate sublimation radius due to the change in luminosity of a 2.9$M_{\odot}$ star, Z=0.02, as it evolves. Objects initially outside the ice-line on the main sequence, e.g. at 100AU, could end up inside of the ice-line around a giant or AGB star, such that any particles composed purely of water ice would sublimate. }
\label{fig:sub}
\end{figure}

\subsubsection{Stellar wind pressure}

Mass loss is an important feature of stellar evolution along the giant, horizontal and asymptotic giant branches, however mass loss rates are low and do not affect the disc significantly except towards the end of the AGB. 
It should be noted that there is a relatively large degree of uncertainty in the exact mass loss rates of an evolving star, as discussed in \S \ref{sec:star}. 
The effect of the stellar wind on particles in the disc is very similar to that of stellar radiation in that its pressure component causes the smallest particles created in collisions to have eccentric or unbound orbits and its drag component causes them to spiral inwards. 
Assuming a stationary wind model, with a constant wind velocity $v_{SW}$, the ratio of the pressure force due to the stellar wind to the gravitational forces is given by:

\begin{equation}
\beta_{SW}= 5.67 \times 10^{10}\frac{\dot M_* v_{SW} Q_{SW} }{M_* \rho D},
\end{equation}
 where $\langle Q_{SW} \rangle$ is the efficiency for momentum transfer from the stellar wind, assumed to be 1, $\dot M_*$ the mass loss rate, in $M_{\odot}$ yr$^{-1}$ and $v_{SW}$ the wind velocity, in kms$^{-1}$.

Just like with radiation pressure grains smaller than $D_{SW}$ or with $\beta_{SW} >0.5$ would be removed from the system, where
\begin{equation}
D_{SW}=1.13\times 10^{-4}\frac{\dot{M_*}v_{SW}}{\rho M_*}\;\; \mathrm{m}.
\label{eq:dsw}
\end{equation}

However, as shown in Fig.~\ref{fig:diam_lim} $D_{SW}$ is smaller than $D_{bl}$ throughout the star's evolution, such that removal of grains by stellar wind pressure can be ignored.

\subsubsection{Stellar wind drag}

The stellar wind causes particles of diameter, D, in $\mu$m, to spiral in towards the star on timescales of: 
\begin {equation}
t_{SW} = 9.4 \times 10^{-17} \frac{ D \rho r^2}{ Q_{SW} \dot M_*} \; \; \; \mathrm{Myr}.
\label{eq:tsw}
\end{equation}

Similarly to for PR-drag these timescales can be compared to those for collisions Eq.~\ref{eq:tc} to derive a condition for the diameter below which particles are removed by stellar wind drag: 

\begin{equation}
D_{SWPR}=194 \frac{\dot{ M_*^2}r^{14/3}\frac{dr}{r}^2Q_D^{* 5/3}D_c}{M_{tot}^2M_*^{8/3}e^{10/3}} \;\; \mathrm{m}.
\label{eq:dswpr}
\end{equation}

In Fig.~\ref{fig:diam_lim} $D_{SWPR}$ is compared to $D_{bl}$, $D_{SW}$ and $D_{PR}$ throughout the star's evolution. It shows that stellar wind drag is only important for the stronger mass loss rates on the horizontal and asymptotic giant branches, seen in Fig.~\ref{fig:star}. This is generally true for all disc and star parameters considered in this study.

The strong mass loss rates, however, only act for a relatively short timescale, shorter than collisional timescales, such that $D_{SWPR}$ actually overestimates the size objects that are removed by stellar wind drag. Fig.~\ref{fig:maxdiamsw} shows the maximum diameter particle that can be removed by stellar wind drag for a disc around a 2.9$M_{\odot}$ star, given the finite AGB lifetime. For the smallest radii discs, since planetesimals up to $D_C$ (1.9km) are present in our model, almost all the mass in the disc is removed by stellar wind drag during the AGB phase. However the maximum diameter particle that can be removed for large radii discs is not much larger than the blow-out size.

 Although the majority of the disc mass at the end of the AGB still lies within the main belt, our treatment of the effects of stellar mass loss do not include the fate of smaller bodies migrating inwards under stellar wind drag. This is the main difference between our work and numerical simulations such as \cite{dong10}. \cite{dong10} also include the effect of planets on a planetesimal belt, including the trapping of planetesimals into mean motion resonances. Although our models do not include the flux from small bodies spiralling in under stellar wind drag in the calculation of the disc luminoisities, the amount of material distributed between the inner edge of the belt and the star has been monitored. This will be discussed further in \S \ref{sec:wdobs}, in terms of the hot white dwarf discs observed around some stars, for example \cite{farihi09}.

\subsubsection{Sublimation}
\label{sec:sub}
As the star evolves to higher luminosities particles heat up and may sublimate. For some ideal assumptions, the resulting rate of change of diameter, D (in m), is independent of the size \citep{jurasmallasteroid}:

\begin{equation}
\label{sub}
{dD\over {dt}} = \frac{2\dot{\sigma}_0}{\rho} \sqrt{\frac{T_0}{T(t)}} e^{ {\frac{ -T_0}{T(t)}}}
\label{eq:sub}
\end{equation}
where T the temperature, in K, $\dot{\sigma}_0=1.5 \times 10^{10}$kgm$^{-2}$s$^{-1}$ and $T_0$ the composition dependent sublimation temperature. For pure water ice $T_0= 5,530$ K and for olivine $T_0=65,300$K, meaning that water ices sublimate at $\sim$110K
, whereas silicates only sublimate at $\sim 1,300$K. Here we define the ice-line and silicate sublimation radius as the radius inside of which black bodies have temperatures hotter than this.

The change in the position of the ice-line and silicate sublimation radius as the star's luminosity changes is shown in Fig.~\ref{fig:sub}. It can be seen that the silicate sublimation radius is always smaller than the discs considered in the current models, hence the sublimation of silicates can be ignored for the population of discs considered.

Temperatures hot enough for the sublimation of water ices, on the other hand, are found in debris discs around main sequence stars, for example sublimation of water ice is important for comets in our solar system on orbits that approach the sun within the ice line of 6 AU. 
 A disc initially outside of the ice-line on the main sequence, may be inside of it by either the giant or asymptotic giant branches. The sublimation of objects composed entirely of water ice would therefore be expected, resulting in significant mass loss from objects of all sizes in the disc. Since sublimation loss timescales are proportional to diameter, this means that smaller objects are removed most rapidly which could truncate the collisional cascade size distribution at a size larger than the blow-out limit. 

However, the behaviour of more realistic objects of mixed composition is more complex. Sublimation may not proceed at the rate given by Eq.~\ref{eq:sub} indefinitely as water ice below the surface may be protected from sublimation by the surrounding layers of other non-volatile material \citep{jura2010}. As observed for solar system comets, sublimation may also lead to the release of small dust grains that were originally embedded in the ice, thus increasing the number of small grains. Such a process was invoked in the models of ~\cite{juraotherkb}. Thus, although sublimation may truncate the size distribution, and so reduce the number of small grains, it may also lead to the production of an extra population of small grains. Due to this complexity in behaviour it is not clear that sublimation cleanly truncates the size distribution and it is therefore assumed not to dominate over other processes in the current models and its effect is discussed further in \S~\ref{sec:giants}.

\subsubsection{Summary}
The five processes that could potentially remove the smallest particles from the disc have been discussed. Fig.~\ref{fig:diam_lim} provides a summary of which processes are relevant as the star evolves. Radiation pressure removes the largest particles from the disc throughout most of its evolution, apart from on the AGB when stellar mass loss rates are high and relatively large objects are removed from the collisional cascade by stellar wind drag. PR-drag is only relevant for small radii discs on the main sequence and giant branch. Uncertainites in the outcome of sublimation mean that the models presented in this paper assume that discs are unaffected by this process; the implications of this assumption are discussedin \S~\ref{sec:giants}. The maximum of $D_{bl}$, $D_{SWPR}$, $D_{PR}$ and $D_{SW}$, as shown in Fig.~\ref{fig:diam_lim}, was used to determine the cut-off of the collisional cascade, $D_{min}$, in our models. For epochs where $D_{min}$ decreases with time, a time delay would be expected before the small grains are replenished by collisions. However in these models we assume that this delay is shorter than the timescales considered and that the collisional cascade is instantaneously replenished.  For example at the start of the white dwarf phase $D_{min}$ decreases rapidly, however small grains will be replenished quickly by collisions. The collisional lifetime for small bodies is short, even though the collisional lifetime for the largest bodies in the disc is long \S ~\ref{sec:mtot}.


\begin{figure}
\includegraphics[width=80mm]{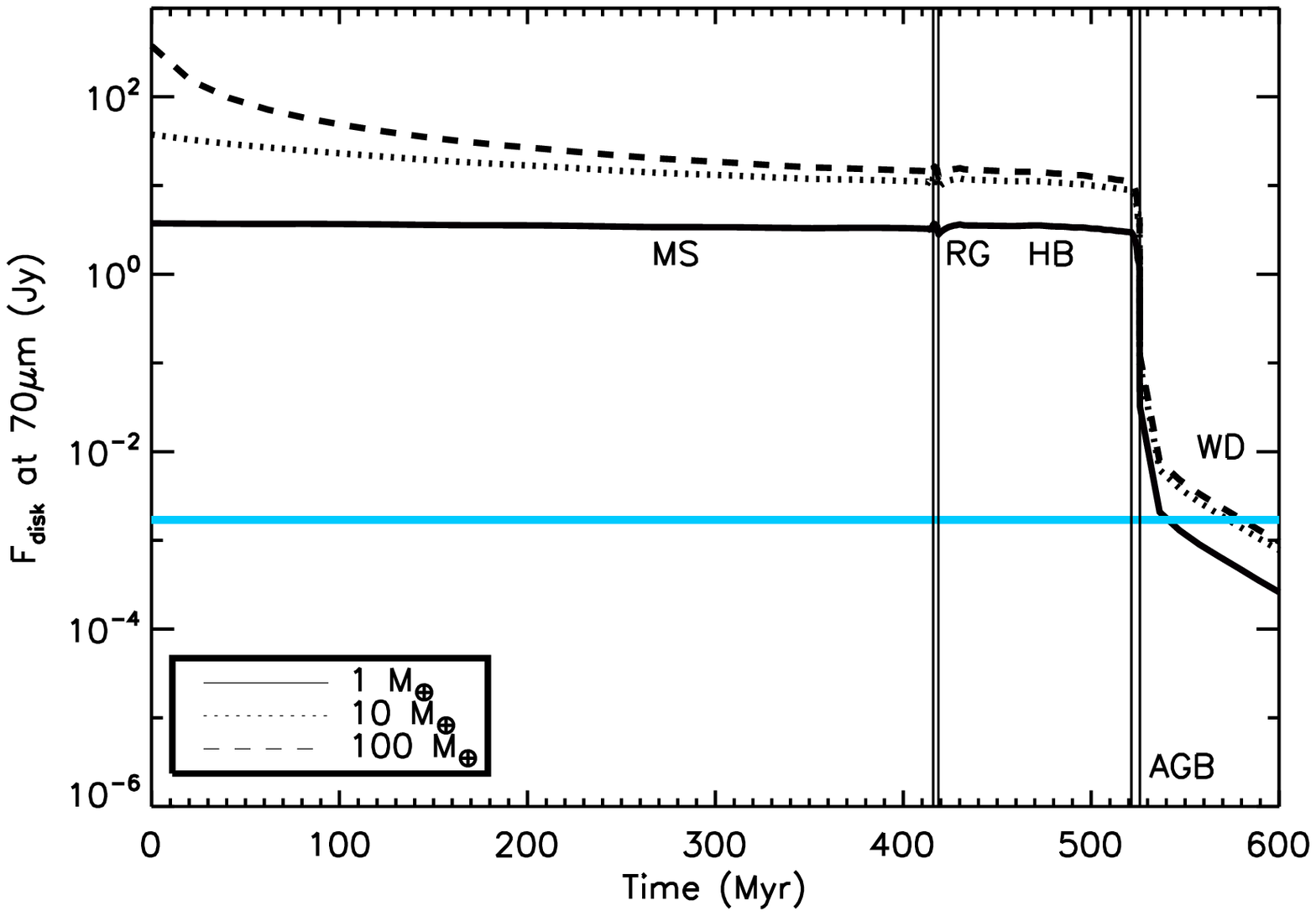}
\includegraphics[width=80mm]{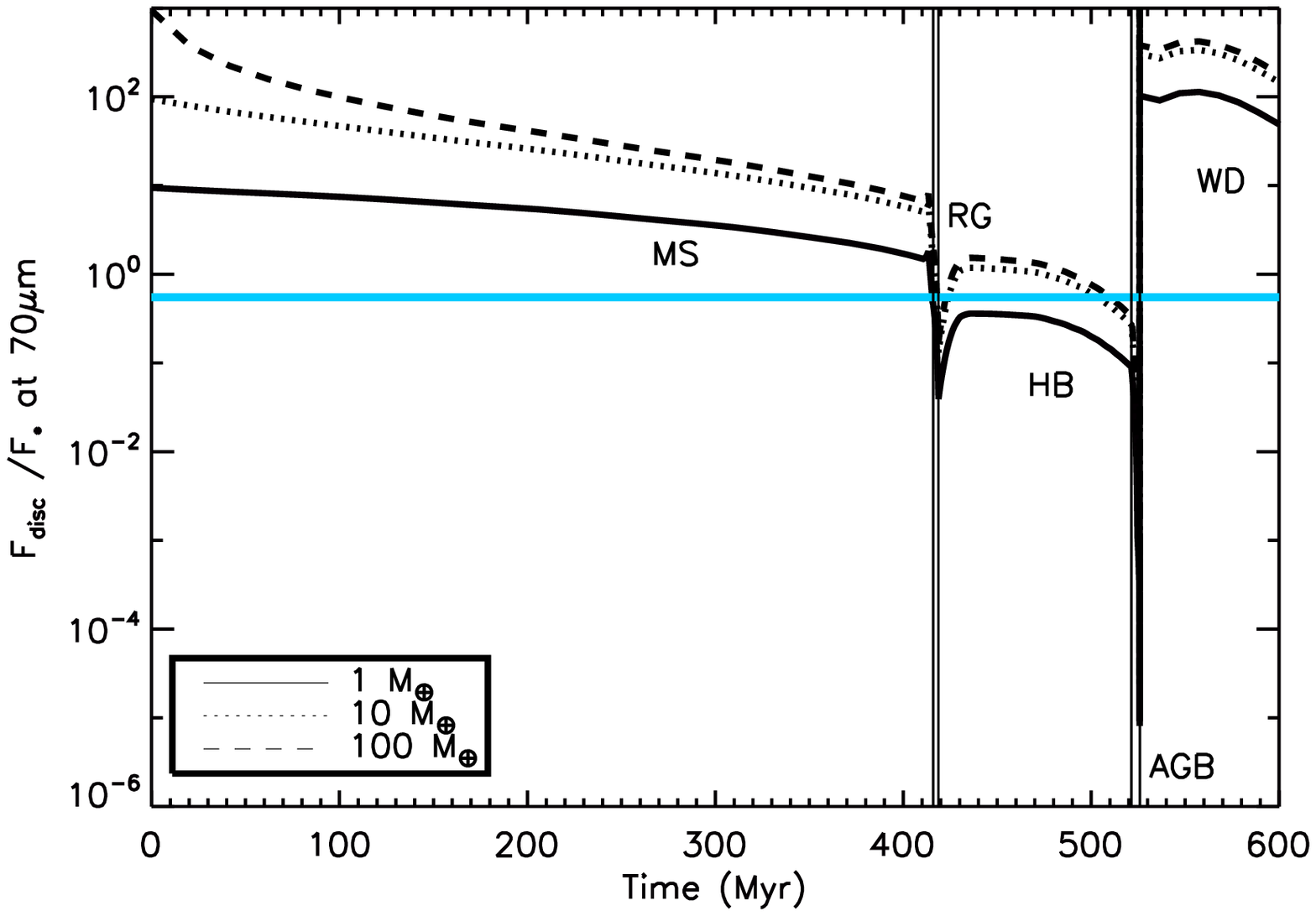}

\caption{The evolution of the total flux from the disc (top panel) and the ratio of the flux from the disc to the flux from the star (middle panel) at $70\mu$m, as the star evolves. The thick blue line in the upper plot is the sensitivity limit of 110$\mu$Jy, whilst in the middle plot it shows the calibration limit of $R_{\mathrm{lim}}=0.1$, for Spitzer at $70\mu$m. The star is a 2.9$M_{\odot}$ star, with solar metallicity ($Z=0.02$), at 10pc and the disc has an initial radius of 100AU.}
\label{fig:f_R}
\end{figure}

\section{Observations}
\label{sec:obs}

The preceding section discussed the various processes that affect the disc as the star evolves. Combining these processes, together with a knowledge of the change in stellar properties as the star evolves (Fig.~\ref{fig:star}), the evolution of a debris disc and its observable properties on the post-main sequence can be modelled. The evolution of an individual disc is, of course, dependent on its initial conditions, characterized in the current models by its radius, initial mass, distance from earth and the spectral type of the star. In this section we first consider the evolution of an individual disc, with a given set of parameters, and then proceed to discuss the evolution of the population of debris discs observed on the main sequence around A stars.

The two quantities of relevance to observations of the disc are its flux ($F_{\mathrm{disc}}$) and the ratio of the flux from the disc to the flux from the star ($R_{\nu}$). In order for a disc to be detected by a given instrument, at a given wavelength, its flux ($F_{\mathrm{disc}}$) must be above the sensitivity limit for that instrument ($F_{\mathrm{sens}}$), and the ratio of the flux from the disc to the flux from the star ($R_{\nu}$) must be above the calibration limit ($R_{\nu,\mathrm{ lim}}$). The calibration limit is set by the accuracy that the stellar flux is known and the quality of the instrumental calibration. Here it is assumed that all far-IR measurements have  the same calibration limit as Spitzer at 70$\mu$m ($\sim$0.55), whilst mid-IR measurements, such as Spitzer at 24$\mu$m have a calibration limit of $\sim$0.1.  Although the instrument calibration for Herschel is quoted as 0.1 \citep{pacs2010}, once the uncertainity in the stellar flux is included the limit will be similar to that for Spitzer at 70$\mu$m.

\subsection{Evolution of a 100AU disc around a 2.9M$_{\odot}$ star at a distance of 10pc}
Fig.~\ref{fig:f_R} shows the evolution of $F_{\mathrm{disc}}$ and $R_{70\mu m}$ at 70$\mu$m, for a disc initially at 100AU, with a mass of 1, 10 or 100$M_\oplus$, around a 2.9$M_{\odot}$ star at 10pc. The blue lines show the sensitivity and calibration limits for Spitzer at 70$\mu$m, respectively. The disc is detectable if both $F_{\mathrm{disc}}$ and $R_{70\mu m}$ are above these limits. From these plots it can be seen that a disc of these initial conditions can be detected on the main sequence, early on the giant branch and early in the white dwarf phase. The exact values of $F_{\mathrm{disc}}$ and $R_{70\mu m}$, relative to the calibration and sensitivity limits, vary significantly for discs of different radius, initial mass, distance from earth or around different spectral type stars, however the form of these plots, in terms of when $F_{\mathrm{disc}}$ and $R_{70\mu m}$ increase or decrease relative to the evolutionary phase of the star, remains relatively unchanged. The discussion begins below by considering the variation of $F_{\mathrm{disc}}$ and $R_{70\mu m}$ during the evolution of a disc with a given set of initial parameters and then goes on to consider the changes to this evolution when these initial parameters of the disc are varied in \S~\ref{sec:discparam}, \S~\ref{sec:starparam} and \S~\ref{sec:wav}.

Along the main sequence the stellar properties change only by a small amount and the evolution of the disc is unchanged from that in \cite{wyatt07}. The flux from the disc falls off with time as collisional evolution depletes the mass in the disc. Observations of discs around nearby stars with Spitzer at $70\mu$m are in general calibration limited, and the example shown at 100AU is detectable throughout the main sequence.

 On the giant branch, the stellar luminosity increases by several orders of magnitude (see Fig.~\ref{fig:star}). The increase in stellar luminosity heats the disc, however the increase in disc flux is small since all the small grains that would dominate the emission are removed by radiation pressure (see middle panel of Fig.~\ref{fig:beta}). There is a substantial decrease in $R_{70\mu m}$ with time along the giant branch, since the increase in stellar flux is large, whilst the increase in $F_{\mathrm{disc}}$ on the giant branch is relatively small. The difficulty in observing discs around giant stars is therefore the calibration limit, as can be seen in Fig.~\ref{fig:f_R} for the example disc for which $R_{70 \mu m}$ is only greater than $R_{70 \mu m,\mathrm{ lim}}$ for the first half of the star's giant branch evolution. 
\par
As the star moves onto the horizontal branch its luminosity decreases from the maximum value on the giant branch, but remains higher than on the main sequence, whilst the stellar temperature remains low (see Fig.~\ref{fig:star}). The combination of these means that the stellar flux is high and $R_{70\mu m}$ is small, less than $R_{70\mu m,\mathrm{ lim}}$, for the 10M$_{\oplus}$ example disc. This is true for the majority of discs in our population. 

\par
As helium in the core is exhausted, the star swells to become an asymptotic giant branch star. It ejects a significant proportion of its mass in a stellar wind and the smallest grains are removed by stellar wind drag ($D_{min}=D_{SWPR}$ Eq.~\ref{eq:dswpr}). The stellar luminosity increases and heats the disc such that $F_{\mathrm{disc}}$ remains high, despite the fact that $D_{min}$ is relatively large. This means that $R_{\nu } < R_{\nu ,\mathrm{ lim}}$ and discs do not have an observable excess from a debris disc. However AGB stars may be surrounded by expanding circumstellar envelopes of material ejected from the star in a stellar wind and emission from these dust shells would be significantly brighter than a debris disc in the infrared or submm. 

 After mass loss ceases, the white dwarf core evolves swiftly to higher effective temperature at constant luminosity, before the stellar luminosity starts to fall as the star cools as a white dwarf. As the stellar luminosity decreases, $R_{\nu }$ increases and it becomes possible to detect emission from this example debris disc. For this short evolutionary epoch the star is defined as a post-AGB or pre-white dwarf. For the purposes of these models we have defined the post-AGB phase as the 0.1Myr before the start of the white dwarf phase.

\par
There is a sharp drop in stellar luminosity as the stellar envelope is ejected and the stellar core is exposed as a white dwarf. This means that the ratio of the stellar luminosity to the disc flux increases significantly and that observations are no longer calibration limited. However the disc flux falls rapidly below the sensitivity limit as the white dwarf cools and it is this limit that determines whether a white dwarf debris disc is detectable. As discussed in \S \ref{sec:radp} even though there is no process to remove small dust created in collisions, the flux from these small grains is small and does not make a white dwarf debris disc detectable. As can be seen in Fig.~\ref{fig:f_R}, $F_{\mathrm{disc}}$ is only greater than $F_{\mathrm{sens}}$ for very young white dwarfs.  

\subsubsection{Dependence on disc parameters}
\label{sec:discparam}
Changes in $F_{\mathrm{disc}}$ and $R_{\nu}$ with initial disc mass and radius are interlinked. Simplistically $F_{\mathrm{disc}}$, and thus $R_{\nu}$, is proportional to disc mass and hence discs that are more massive are easier to observe. The collisional evolution of material in the disc, however, means that there is a dependence of disc mass at later times on disc radius, since the collisional lifetime is shorter for close-in discs than for those further out (Eq.~\ref{eq:tc}). For discs that have reached collisional equilibrium their mass, at a given age, is independent of their initial mass but increases with disc radius (Eq.~\ref{eq:mmax}). Discs at large radii, on the other hand, will not have reached collisional equilibrium and so retain their initial mass. This leads to the behaviour of $F_{\mathrm{disc}}$ with radius shown in the upper panel of Fig.~\ref{fig:obs}. For close-in discs the disc mass and thus $F_{\mathrm{disc}}$ increases with radius, despite the decrease in disc temperature. For large radii discs, on the other hand, there is a significant variation in $F_{\mathrm{disc}}$ with initial disc mass and $F_{\mathrm{disc}}$ decreases with radius or disc temperature. This behaviour is of particular importance in determining which radii discs are the brightest at a given epoch. As can be seen for a given individual disc mass, the brightest discs at some later epoch are those with intermediate radii at which the largest planestessimals are just reaching collisional equilibrium at this age.

\subsubsection{Dependence on stellar parameters}
\label{sec:starparam}
The simplest scaling relation is the distance to the star. $R_{\nu}$ is unchanged, whilst $F_{\mathrm{disc}}$ scales inversely with distance squared. At large enough distances observations are always sensitivity limited.

In the current models stars with mass between 1.67 and 3.8 $M_{\odot}$ or spectral type A9-B8 are considered. The difference between these models that has the greatest effect on the disc is that in stellar luminosity. More luminous stars have brighter discs, although this increase in not as large as might be expected because the blow-out size also increases with stellar luminosity. Thus, the ratio of the disc flux to the stellar flux decreases with stellar luminosity, since the increase in stellar flux is larger than the increase in disc flux. In terms of Fig.~\ref{fig:f_R} this means that, for higher luminosity stars, the upper plot is shifted upwards relative to the sensitivity limit, whilst the lower plot shifts downwards. As was discussed in \S~\ref{sec:star} the stellar luminosity increases the most on the giant branch for lower mass stars and therefore $F_{\mathrm{disc}}$ and $R_{\nu}$ show the greatest variation on the giant and asymptotic giant branch for these stars.

There is also a dependence in disc mass on main sequence lifetime for collisionally evolving discs. Later spectral type stars take much longer to evolve and therefore the reduction in their disc mass at a given epoch is larger, however the difference in disc flux due to this is small compared to the difference due to the change in stellar luminosity.

 In terms of detecting discs, this means that where observations are calibration limited, disc are more detectable around the least luminous stars. This applies to lower mass stars on the main sequence, or early on the giant branch. On the otherhand when observation are sensitivity limited, i.e. around white dwarfs, discs are more detectable around the most luminous, or higher mass stars.

\subsubsection{Dependence on wavelength of observations}
\label{sec:wav}

\begin{figure}
\includegraphics[width=70mm]{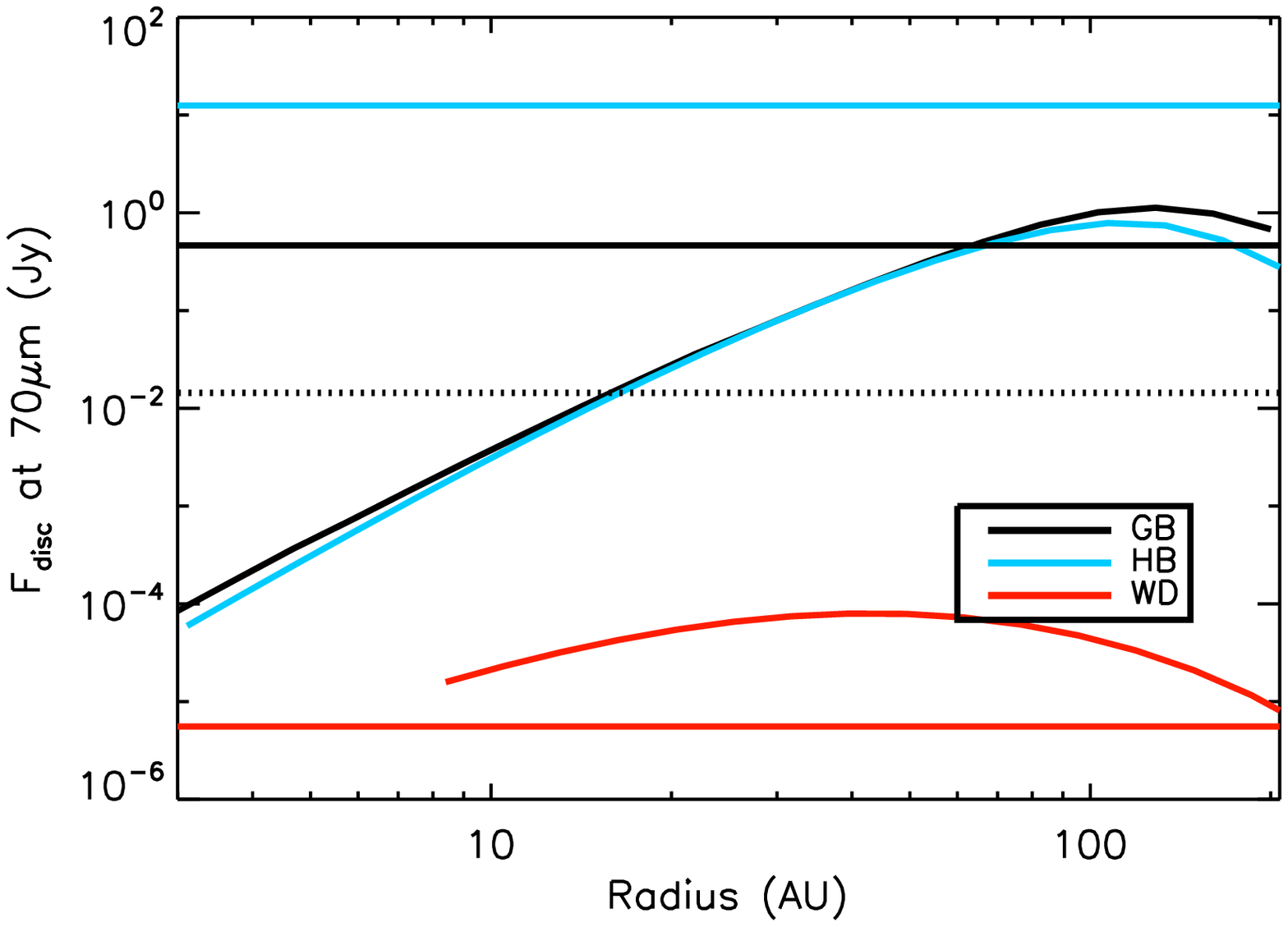}
\includegraphics[width=70mm]{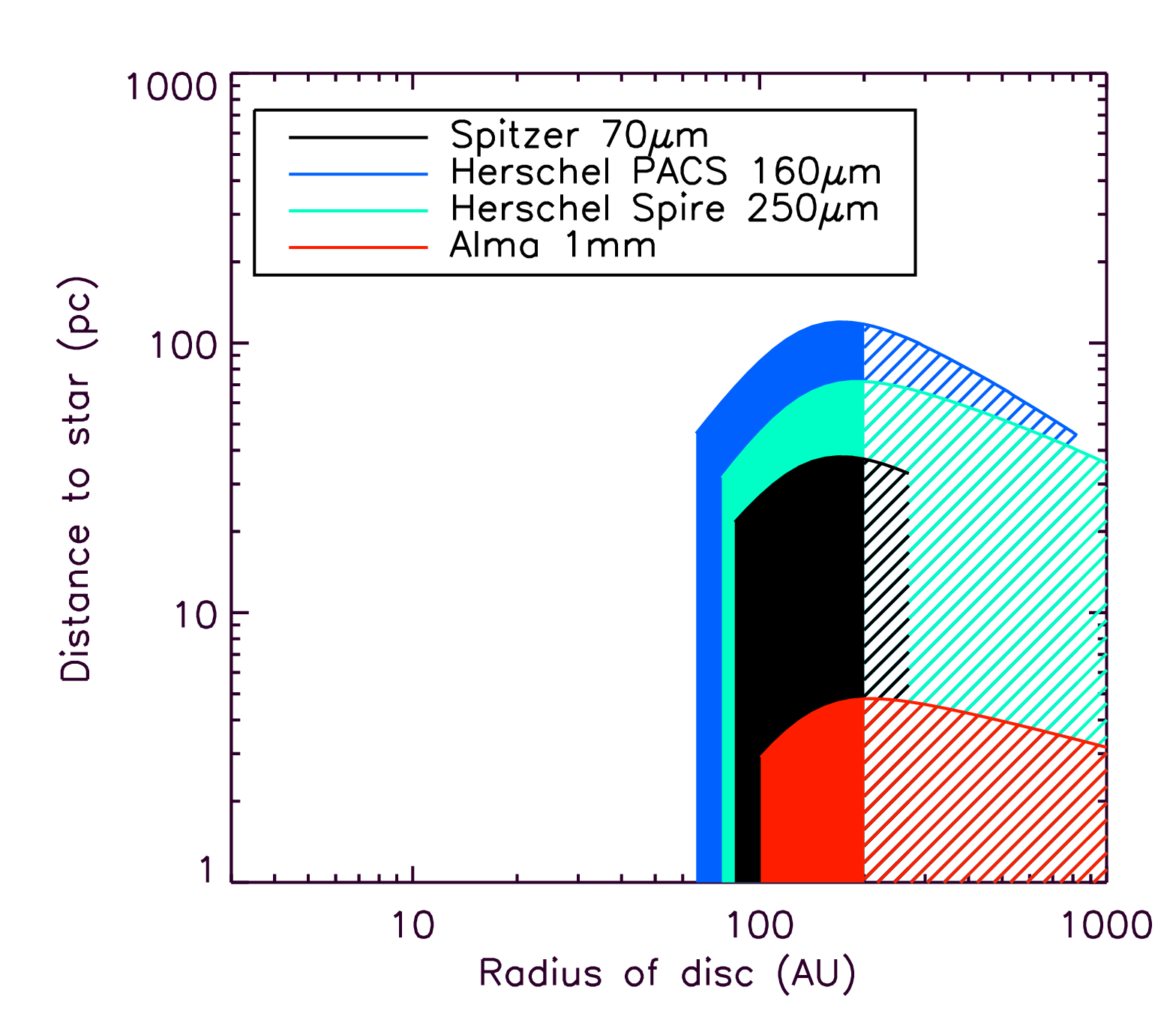}
\includegraphics[width=70mm]{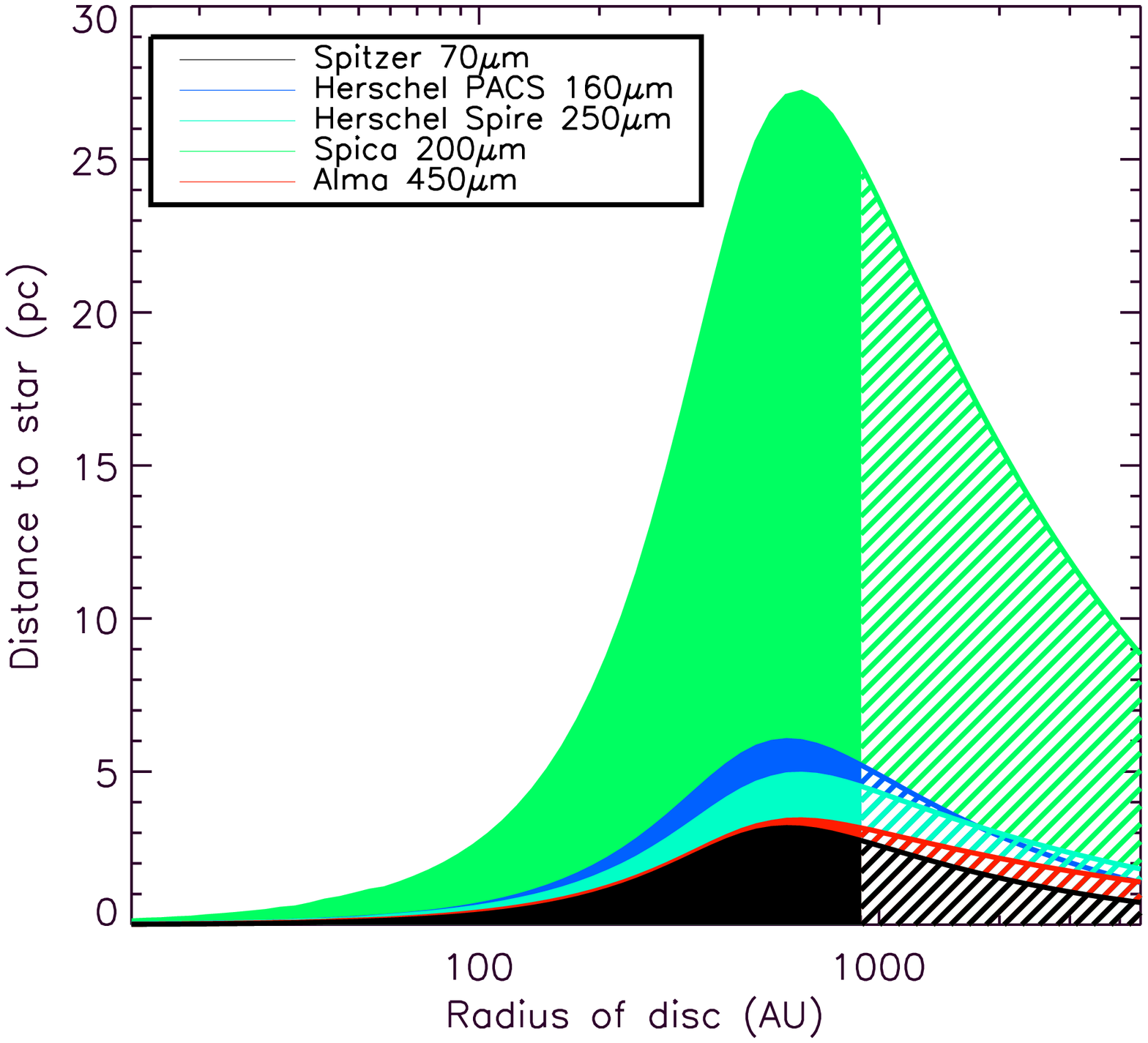}

\caption{Observations of the disc. 
Upper: The variation in disc flux with radius for a 1.67$M_{\odot}$ star on the giant branch (2000Myr) (black line), horizontal branch (2200Myr) (blue line) and around a white dwarf (cooling age of 1Gyr) (red line), for discs with an initial mass of 10$M_{\oplus}$. The horizontal lines show the calibration limits for Spitzer at70$\mu$m on the giant branch (black), horizontal branch (blue) and white dwarf phase (red), respectively, whilst the dotted horizontal line shows the sensitivity limit. 
Middle: Detection limits for discs of initially 10$M_{\oplus}$, around a star of 1.67$M_{\odot}$, at the base of the giant branch ($L_* =10.2 L_{\odot}$). Excesses can be observed for the discs that fall within the solid filled regions for Herschel PACS at 160$\mu$m, Herschel Spire at 250$\mu$m, Alma at 1mm and Spitzer at 70$\mu$m. The dashed filled regions are for discs with radii larger than 200AU, not included in the current models, that can be detected. A calibration limit of $R_{\nu, lim}=0.55$ is assumed for all instruments.
Lower: same as middle but for a 3.8$M_{\odot}$ (equivalent to B8) star that has evolved to become a 1Myr old white dwarf. Note that the disc radii are 3.93 times larger than on the main sequence. }
\label{fig:obs}

\end{figure}

The above discussion has focused on observations with Spitzer at 70$\mu$m. The form of the upper and middle panel of Fig.~\ref{fig:obs} remain relatively unchanged as observations are made in different wavelengths however the exact values of $F_{\mathrm{disc}}$ and $R_{\nu}$ relative to the sensitivity and calibration limits vary significantly. The disc flux peaks at approximately the peak emission wavelength for a blackbody of the disc temperature. The ratio of the disc to stellar flux also has a similar variation with wavelength, however it peaks at longer wavelengths, since $F_*$ falls off more rapidly with wavelength than $F_{\mathrm{disc}}$. Variations in $F_{\mathrm{disc}}$ and $R_{\nu}$ are larger for shorter wavelengths, where the emission is from the Wien region of the blackbody spectrum. 

All of this behaviour, means that there will be an optimum wavelength for detecting discs that depends on disc temperature, and whether observations are sensitivity or calibration limited. When observations are sensitivity limited discs are most detectable for the wavelength at which $F_{\mathrm{disc}}$ is maximum, given by Wien's displacement law for a disc of a given temperature, around 100$\mu$m for young white dwarfs. Alternatively when observations are calibration limited, the most discs are detectable for the wavelengths at which $R_{\nu}$ is maximum, for example on the giant branch this varies between 100 and 800$\mu$m.

\subsection{Population models}
\label{sec:pop}
Using our models the evolution of a disc, with a given set of initial parameters, can be determined. Here we apply these models, to the population of discs on the main sequence known from observations of A stars by Spitzer and the models of \cite{wyatt07} (see \S.~\ref{sec:modelmw}). These discs are evolved from the main sequence through to the white dwarf phase and the population of discs around evolved stars is determined. The following discussion focuses on giant stars, horizontal branch, post-AGB stars and white dwarfs. AGB stars are not discussed because debris discs are not detectable during this phase, and in any case observations would be complicated by the presence of material emitted in the stellar wind. 

There are many surveys for debris discs with recent and current instruments, such as Spitzer and Herschel, as well as up-coming instruments such as Alma. Here we calculate the percentage of the evolved population that are detectable with various instruments. Table~\ref{tab:percent} shows these percentages for young white dwarfs, horizontal branch, giant and main sequence stars.

It is important to note that these percentages only correspond to the population of evolved A stars, not the entire population of giants, horizontal branch stars or white dwarfs. The number of discs that these percentages correspond to can be calculated from the space density of A stars from \cite{phillips09} of $0.0014\pm0.0001$ pc$^{-3}$ and the average main sequence lifetime for A stars, 950Myr, to give a density of $3.5 \times 10^{-5}$pc$^{-3}$ for evolved A stars on the giant branch, $2.5 \times 10^{-4}$pc$^{-3}$ on the horizontal branch and $ 1.47\times 10^{-4}$pc$^{-3}$ for white dwarfs with a cooling age of less than 100Myr. It is not possible to tell from observations of giant stars whether they are evolved FGK or A stars and it is therefore hard to compare the populations, however the majority of white dwarfs currently observed are evolved A stars and these space densities make a reasonable comparison with the 80\% complete catalogue of \cite{wdcat09}; the number of white dwarfs within 10pc less than 100 Myr old is predicted to be 0.6 and less than 1000Myr within 20pc is predicted to be 50, compared to 0 and 30 \citep{wdcat09}. This catalogue contains no white dwarfs with cooling ages of less than 1Myr with 20pc.

\begin{table*}
\centering
\begin{minipage}{\textwidth}
\centering                          
\begin{tabular}{c c rrrrr}          
\hline\hline                        
&  &Main Sequence &  Giant Branch&  Horizontal Branch& Post-AGB&  White Dwarf  \\ [0.5ex]
Instruments& Sensitivity  &  d$<100$pc&  d$< 100$pc &d$< 50$pc & d$< 200$pc &   $d<10$pc\\
& (mJy) & &  &L$<100L_{\odot}$ & $t_{WD}$-0.1Myr$<t<t_{WD}$&$t_{WD}<1000$Myr   \\

\hline                               
& &  $\%$  &  $\%$ &   $\%$ &   $\%$   \\

\hline 

IRAS at 60$\mu$m$^{b}$ &100$^c$& 4.6 &1.7&0.6&1.8&$<$0.1 \\
Spitzer at 24$\mu$m& 0.11 $^d$ &51.0 & 14.0 &20.5& $<$1.0&  $<$0.1 \\
Spitzer at 70$\mu$m&14.4$^d$  &39.0&9.3&$<$1.0&6.3&$<$0.1 \\
Spitzer at 160$\mu$m & 40 $^d$   &13.0&4.2&1.0&5.7&$<$0.1 \\
Herschel PACS at 70$\mu$m & 4$^e$  &44.0&9.6&$<$1.0&9.1&$<$0.1 \\
Herschel PACS at 160$\mu$m & 4$^e$   &35.0&12.2&1.0&9.1&1.6 \\
Herschel SPIRE at 250$\mu$m & 1.8$^e$  &33.0&12.8&2.5&22.6&1.9 \\
Herschel SPIRE at 350$\mu$m & 2.2$^e$  &23.0&10.8&3.0&11.4&1.0 \\
Alma at 450$\mu$m  & 80$^f$ &19.0&7.0&3.0&12.6&1.1 \\
Alma at 1.2mm &0.25& 10.7&2.2&$<$0.1&22.6&2.5 \\
Spica at 200$\mu$m & 0.1$^g$  &45.0&12.0&1.5&22.6&23.70 \\

\hline
No. of stars$^h$ &  & 5860$^b$ & 1050$^b$&130$^b$ &5.0& 6.6$^b$ \\
 
\hline   \hline

\footnotetext[1] {The calibration limit of Spitzer at 70$\mu$m. }
\footnotetext[2]{Only stars with magnitudes brighter than 4.0 are considered such that the sample can be compared with ~\cite{jura90}}
\footnotetext[3]{http://irsa.ipac.caltech.edu/IRASdocs/iras\_mission.html}
\footnotetext[4]{ \cite{wyattreview}} 
\footnotetext[5]{http://herschel.esac.esa.int/science\_instruments.shtml } 
\footnotetext[6]{http://www.eso.org/sci/facilities/alma/observing/specifications/}
\footnotetext[7]{ \cite{spica}}
\footnotetext[8]{The number of evolved A stars, calculated from the space density of A stars \citep{phillips09}}

\end{tabular}
\end{minipage}
\caption{Detection of discs around evolved stars}
\label{tab:percent}
\end{table*}

\subsubsection{Giant stars}
\label{sec:giants}

Early on the giant branch a small set of the evolved population of debris discs have a detectable excess. The following discussion defines which discs are detectable, in terms of the parameter space specified by initial disc radius, initial disc mass, distance to the star, wavelength for observations and mass of the star. In order to assess this the disc flux is plotted as a function of radius in the upper panel of Fig.~\ref{fig:obs}. As discussed in \S~\ref{sec:discparam}, this peaks at intermediate radii because collisions have depleted the mass in close-in discs, such that $F_{disc} \propto M_{max} \propto r^{7/3}$ (Eq.~\ref{eq:mmax}), whereas large disc radii retain their initial masses, and the disc flux falls off with the disc temperature or radius. Only discs with $R_{\nu} > R_{lim}$ are detectable, or those with fluxes above the solid lines in the upper panel of Fig.~\ref{fig:obs}, excluding both small and large radii discs. 

This dependence leads to the form of the middle panel of Fig.~\ref{fig:obs}, the solid area of which shows the discs that can be detected with various instruments. The upper curve is the sensitivity limit, whilst the cut-off at low and high radii are from the calibration limit. This plot varies with mass and age of the star, as well as mass of the disc and wavelength for observations. As can be seen for the example disc shown, around an evolved 1.67M$_{\odot}$ star, of initially 10M$_{\oplus}$ at the start of the giant branch, only discs with radii of around 100AU, within $\sim200$pc of the sun, are detectable with Spitzer, Herschel or Alma at the wavelengths considered. As the luminosity of the star increases along the giant branch, the distance out to which discs can be detected increases, however the range of radii of discs with detectable excess decreases. This means that the solid (detectable) area of an equivalent to the middle panel of Fig.~\ref{fig:obs} is largest for the least luminous giants. The dependence of disc flux on wavelength discussed in \S~\ref{sec:wav} means that the solid (detectable) area is largest for Herchel PACS at 160$\mu$m.

A smaller fraction of the population has detectable excess on the giant branch than the main sequence, as can be seen in Table~\ref{tab:percent}. The significant increase in stellar luminosity, compared to the small increase in disc flux means that fewer discs are detectable over the stellar emission ($R_{\nu} > R_{lim}$). Spitzer at 24$\mu$m can detect the largest fraction of the population, because observations are calibration limited and the calibration limit in the mid-IR is lower than in the far-IR. Herschel SPIRE detects the next highest fraction of the population due to the wavelength dependence of $R_{\nu}$, peaking in the sub-mm, as discussed in \S~\ref{sec:wav}. A sample of stars within 100pc were considered in Table.~\ref{tab:percent}. However, if observations with, for example, Alma were made with the intention of detecting such discs, a sample that only extended out to smaller distances would maximise the rate of detection.

Our models suggest that around 10\% of evolved A stars on the giant branch have detectable excess with Spitzer or Herschel. This is, however, subject to the unclear effect of sublimation on debris discs. Sublimation could have two possible effects. Either it removes all small grains, truncating the collisional cascade, and thus decreasing the number of discs with detectable excess, or it releases a population of small silicate grains, increasing the number of giants with detectable excess. This makes future observations of giant stars with Herschel, in comparision with our models, very interesting, as they have the potential to constrain the effects of sublimation on discs.

Our models, however, compare favourably with the sample of 44 giants brighter than $m_v=4.0$ mag \citep{jura90} (see Table~\ref{tab:percent}), none of which display excess at 60$\mu$m ($<3\%$) with IRAS. Infrared excess is, however, found around 300 of the 40,000 G and K giants in the Bright Star Catalogue and Mitchigan Spectral Catalogue \citep{zuckermankim95}, although the origin of this emission is not clear. 12 of these sources are modelled in further detail in \citep{kimzuckerman01}, who suggest that they are more likely to result from sporadic dust ejection or emission from nearby interstellar cirrus rather than black-body grains in a Kuiper-belt disc. In order to compare observations of giant stars with our models it would be necessary to distinguish between these scenarios, potentially with high resolution imaging.

 Another factor that could significantly change the detectability of discs around giant stars is the radius distribution of discs in our population. Our models only included discs detected with Spitzer at 24$\mu$m and 70$\mu$m and therefore there is a bias towards small radii discs. { \it This is particularly relevant, as large radii discs are detectable, particularly at longer wavelengths (see middle panel of Fig.~\ref{fig:obs}).} This could be accounted for by extending our models to include sub-mm observations of debris discs on the main sequence or incorporating this radius bias into our modelling of the main sequence population. Including observations of discs around main sequence FGK stars would also make our models more directly comparable with a sample of giant stars. Given these extensions and a technique to distinguish emission from a debris discs from stars undergoing sporadic dust ejection or emission from nearby interstellar cirrus, it should be possible for future observations with Herschel or Alma to determine the effect of sublimation on debris discs.

\subsubsection{Horizontal branch stars}
The majority of the discussion in \S~\ref{sec:giants} also applies to horizontal branch stars. Observations are also calibration limited, however significantly fewer discs are detectable around horizontal branch stars than giant stars, since the stellar flux is on average higher, whilst the disc flux remains approximately constant. In order to maximise the percentage of discs with detectable excess observations should focus on low luminosity horizontal branch stars. In Table~\ref{tab:percent} the percentage of the population of horizontal branch stars within 20pc and with luminosities lower than 100$L_{\odot}$ were calculated. The most discs are detectable with Herschel PACs at 160$\mu$m since this is the wavelength at which $R_{\nu}$ is maximum. 

\subsubsection{White dwarfs}
\label{sec:wdobs}
As can be seen in Fig.~\ref{fig:f_R} the disc flux falls off rapidly as the star cools during the white dwarf phase and it is therefore very hard to detect debris discs around white dwarfs. Observations of debris discs around white dwarfs in our baseline model are sensitivity limited and only the most massive discs around the closest, youngest white dwarfs are detectable. By the same analysis as in \S~\ref{sec:giants} the bottom panel of Fig.~\ref{fig:obs} shows the distance out to which discs of initially $10$M$_\oplus$ can be detected around an evolved 3.8$M_{\odot}$ (equivalent to B8) star, with a white dwarf cooling age of 1Myr. Thus for discs in our baseline model (initial radii less than 200AU) around white dwarfs that are younger than 1 Myr, the disc flux is so low that it is only those that are within a couple of parsecs of the Sun that are detectable with Spitzer, Herschel or Alma. Even the increased sensitivity of Spica only means that discs out to tens of parsec are detectable.

Similarly to discs around giant stars, it is the large radii discs that retain the highest mass at late times, that are therefore the most detectable. As can be seen in the upper panel of Fig.~\ref{fig:obs}, $F_{\mathrm{disc}}$ peaks at $\sim$200AU for a disc of initially 1M$_{\oplus}$ around the 1Myr old white dwarf considered. This radius increases with initial disc mass or white dwarf cooling age. 

\begin{figure}
\includegraphics[width=80mm]{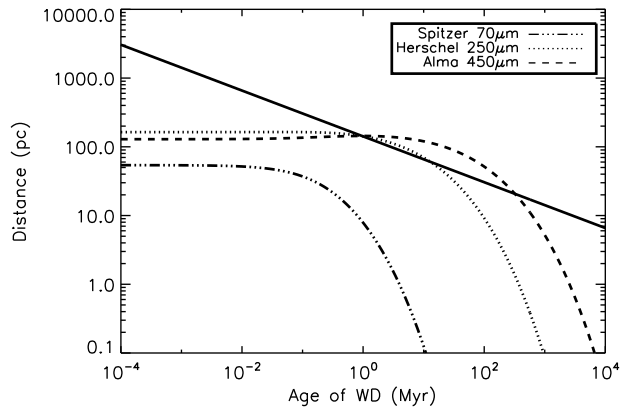}
\caption{The solid line shows the maximum distance out to which a disc initially at 100AU, with a mass of $10M_{\oplus}$, around an evolved 3.8$M_{\odot}$ white dwarf, of a given age can be detected with Spitzer at 70$\mu$m, Herschel at 250$\mu$m and Alma at 450$\mu$m, whilst the dotted line shows the distance within which there is one white dwarf younger than the given age, calculated using the space density of A stars from \citet{phillips09}. }
\label{fig:wdmaxdist}
\end{figure}

There is, however, a balance between young white dwarfs being the most luminous and therefore having the brightest discs and the low volume density of young white dwarfs such that they are more likely to be found at greater distances from the sun. Fig.~\ref{fig:wdmaxdist} shows the maximum distance out to which discs around white dwarfs can be detected as a function of cooling age, for discs at 100AU with Spitzer at 70$\mu$m, Herschel SPIRE at 250$\mu$m and Alma at 450$\mu$m. This is compared to the distance within which one white dwarf of a given cooling age is found, according to the space densities of \cite{phillips09}. The maximum distance out to which discs can be detected is never significantly greater than the distance within which there is one white dwarf and it is therefore unlikely that such a system can be observed. There is an optimum cooling age for detecting white dwarf discs, which varies with wavelength, for Spitzer at 70$\mu$m it is $\sim$1Myr, whilst for Herschel SPIRE at 250$\mu$m it is $\sim$10Myr and Alma at 450$\mu$m $\sim$100Myr. As the disc temperature drops, the disc flux decreases, more rapidly at the shorter wavelengths. This means that for a young population of white dwarfs, the best chances of detecting debris discs are at the shorter wavelengths of Spitzer or Herschel, whilst for a sample that includes older stars Alma would be better. However, overall, the best chances of detecting such a system are with the longer wavelengths of Herschel or Alma.

Focusing on Spitzer at 70$\mu$m, if for some reason our models under-predicted the flux from (or mass in) such discs by approximately an order of magnitude a disc would be most likely to be detected around a white dwarf of less than 5Myr old at a distance of around 200pc. The only detection of excess around a white dwarf that resembles a main sequence debris disc is the helix nebula \citep{helix}, a young white dwarf with a cooling age significantly less than 5Myr, surrounded by a planetary nebula at 219pc. This fits nicely with our models, especially given that alternative explanations that increase the disc flux exist, for example the trapping of bodies in resonances \citep{dong10}.

There are very few young white dwarfs close to the sun, therefore assuming that our models are correct, the best chances of detecting a white dwarf debris disc are to observe nearby white dwarfs with Alma. Table~\ref{tab:percent} shows the percentage of the population of white dwarfs within 10pc with a cooling age of less than 1000Myr. According to the space densities of \citep{phillips09} there are only ~6 white dwarfs in this distance, and even less from \cite{wdcat09}, and therefore the chances of one of these white dwarfs having a debris disc within the narrow initial radius and initial mass range such that it is detectable is slim. Increasing the distance limit of the sample does not improve matters as at greater distances the discs flux falls below the sensitivity limit. Even with the increased sensitivity of Spica the chances of observing such a disc around a white dwarf are slim.

\begin{figure}
\includegraphics[width=80mm]{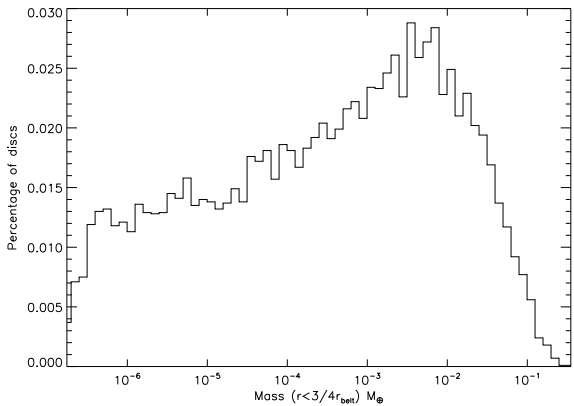}
\caption{A histogram showing the amount of mass left inside of the main belt (between r=0 and $r=\frac{3}{4} r_{belt}$) at the end of the AGB for the population of discs in our models. }
\label{fig:hist}
\end{figure}

These low probabilities for detecting debris discs around white dwarfs fits with the fact that Spitzer observations of white dwarfs that have only found one white dwarf with infrared excess fitted by a disc with a radius of the same order of magnitude of main sequence debris discs. There are however $\sim$20 observations of hot, dusty discs around white dwarfs that are best fitted by discs of radii on the order the solar radius e.g. \cite{farihi09}, \cite{reach05}.  \cite{farihi09} estimate that 1-3$\%$ of white dwarfs with cooling ages less than 0.5 Gyr possess hot IR excess. The minimum radius of a disc in our population is $\sim$10AU and therefore these observations cannot be explained by the discs in our population. Material in discs with such a small radius will have a very short lifetime and must, therefore, be replenished. Within the context of the current model we have identified a potential source of material for such discs. Stellar wind drag was included in the current models in as far as it truncates the collisional cascade on the AGB. Material that leaves the disc will spiral in towards the sun, most of it being accreted onto the star during the AGB, however some mass will be left between the inner edge of the belt and the star, at the end of the AGB. Fig.~\ref{fig:hist} shows this mass for all the discs in our population. The masses in Fig.~\ref{fig:hist} are significantly higher than the typical dust masses for these hot discs e.g $3.3 \times 10^{-10} M_\oplus$ of GD166-58 \citep{farihiI}, and there are even a significant proportion of the population with more mass than the largest such disc, GD362, with a mass of 0.017$M_{\oplus}$ \citep{juraxray}. However a mechanism is still required to move this material in closer to the star. This could potentially be scattering by planets inside of the disc or the dynamical effects of mass loss on the disc. The effects will be considered in more detail in a future piece of work.

 \subsubsection{Post-AGB or pre-WD stars}
As discussed earlier in Sec~\ref{sec:starparam}, the stellar flux from AGB stars is so high that it is hard to detect emission from a debris disc. However as the stellar luminosity starts to drop just before the start of the white dwarf phase it is possible to detect discs around a small proportion of stars (see Table.~\ref{tab:percent}). The analysis is very similar to that for young white dwarf stars, discussed in the previous section. The only difference is that it is possible to observe discs around somewhat more distant stars, however since the post-AGB phase is shorter the density of such stars is lower. The chances of detecting such a system are therefore slim. 
 Many post-AGB stars have IR excess that can be modelled as a stable Keplerian disc (see \cite{post-agbreview} for a review). However these discs are orders of magnitude brighter than a debris disc at this epoch and are most likely to be connected with binarity \citep{vanwinckelbinarypostagb}.

\section{Conclusions}
\label{sec:conclusions}

This work provides a theoretical framework that considers all of the effects of stellar evolution on debris discs, firstly considering the evolution of an individual disc, before extending this to the known population of debris discs around main sequence A stars. It is found that debris discs are harder to detect around evolved stars than on the main sequence. The fraction of discs with detectable excess decreases significantly on the giant branch, yet further on the horizontal branch and discs around white dwarfs are very hard to detect, although the limitations during this phase are different to earlier in the star's evolution.

The population of discs on the main sequence is constrained by Spitzer observations of A stars \citep{rieke05, su06} and the steady state collisional models of \cite{wyatt07}. In this work these models were updated to include realistic grains rather than the black body approximation used in \cite{wyatt07}. This was done relatively simplistically by considering that the difference in behaviour between realistic and black body grains can be explained entirely by an altered disc radius, characterised by the ratio $X_{2470}$ between the radius calculated using realistic grains (r) and the radius calculated using a black body approximation (r$_{2470}$), shown in Fig.~\ref{fig:ratio}. This was used to adjust the fit from \cite{wyatt07} and thus the population of discs around main sequence A stars was determined.

In our models debris discs that are observed on the main sequence survive the star's evolution, however their properties are altered. They evolve collisionally in exactly the same manner as on the main sequence, however the longer timescales mean that, except for large radii discs, their masses are significantly reduced. Discs heat up as the stellar luminosity increases on the giant and asymptotic giant branches. The increase in disc flux, however, is relatively small since small grains are removed by radiation pressure and stellar wind drag (on the AGB only). It is shown that Poynting-Robertson drag is irrelevant for all discs, including discs around white dwarfs, the only exception being for low mass or close-in discs. Adiabatic stellar mass loss means that discs around white dwarfs have radii a factor of 2 or 3 greater than on the main sequence. 

All of these changes in the properties of the disc can be put together to determine which discs can be detected. In terms of observations of discs around post-main sequence stars, the important quantities are the disc flux $F_{\mathrm{disc}}$ and its ratio to the stellar flux, $R_{\nu}$, which must, respectively, be above the sensitivity and calibration limits of the instrument considered. The variation in these are summarised in Fig.~\ref{fig:f_R}.

A smaller fraction of the population can be detected on the giant branch than the main sequence. Observations are calibration limited. The large increase in stellar flux at the wavelengths considered, compared to the smaller increase in disc flux, means that $R_{\nu}$ decreases significantly and discs are hard to observe. $R_{\nu}$ decreases with time on the giant branch such that only large radii discs around stars early on the giant branch have a detectable excess. One limitation of our models is the uncertainty in the effect of sublimation on the disc, as discussed in \S \ref{sec:sub}. Future observations of giant stars with Herschel or Alma, in comparison with our models, could potentially constrain the effects of sublimation on debris discs.

Discs around white dwarfs are very faint and thus hard to observe. Their luminosity decreases as the stellar luminosity falls off with age and the best chances of observing a disc are around very young white dwarfs close to the sun, however there are very few such objects and thus the chances of observing such a system are small. If for some reason our models underpredict the flux from such discs, then the optimum age and distance for detecting a white dwarf disc with Spitzer at 70$\mu$m would be at a distance of $\sim$200pc and an age of less than 5Myr. This fits nicely with the one detected disc around WD2226-210, the young white dwarf at the centre of the helix nebula, at a distance of 219pc \citep{helix}.

There are however detections of hot dusty discs around $\sim$20 white dwarfs with radii less than 0.01AU. These are a different population from the discs in our models which have a minimum radius of 6AU.  Within the context of our models a potential source of material to replenish such discs is material that has left the disc due to stellar wind drag, but not yet been accreted onto the star at the end of the AGB. However, a mechanism is still required to move this material in towards the star. This, and the dynamical effects of mass loss on debris discs will be the subject of future work.

\bibliographystyle{mn.bst}

\bibliography{ref}

\end{document}